\documentclass[12pt]{article}
\title{\bf Representations of coefficients of power series in classical 
statistical mechanics.
Their classification and complexity criteria.
}
\author{\bf G.I.~Kalmykov}
\date{}
\usepackage{amsthm,latexsym}
\usepackage{amsfonts}
\usepackage{amssymb}
\usepackage{amsmath}
\usepackage{bm}
\usepackage[cp1251]{inputenc}
\usepackage[T1,T2A]{fontenc}
\usepackage[russian,english]{babel}
\usepackage{indentfirst}
\usepackage{enumerate}
\usepackage{verbatim}
\newcommand{\MO}{\mathop}
\newcommand{\LT}{\lefteqn}
\newcommand{\R}{{\bf R}^{\nu}}
\newcommand{\XX}{\widetilde X}

\renewcommand{\refname}{Список литературы}
\begin{document}
\maketitle

\begin{abstract}
It is declared that the aim of simplifying representations of coefficients
of power series of classical statistical mechanics is to simplify
a process of obtaining estimates of the coefficients using their simplified
representations.

The aim of the article is: to formulate criteria for the complexity (from
the above point of view) of representations of coefficients of the power series
of classical statistical mechanics and to demonstrate their application
by examples of comparing the Ree-Hoover representations of virial coefficients
(briefly --- the Ree-Hoover representations) with such representations of power
series coefficients that are based on the conception of the frame classification
of labeled graphs.

To solve these problems, mathematical notions were introduced (such as 
the basic product, the basic integral, the basic linear combination 
of integrals, the basic linear combination of integrals with coefficients 
of insignificant complexity(the abbreviation --- BLC with CIC) 
and the classification of representations of the coefficients of
power series of classical statistical mechanics is proposed. 
In the classification, the class of BLCs with CIC is the most important. 
It includes the most well-known representations of the coefficients 
of power series of classical statistical mechanics.

Three criteria are formulated to assess the comparative complexity 
of BLCs with CIC and their extensions to the set of polynomials
from BLCs with CIC are constructed.
The application of all the constructed criteria is demonstrated by examples
of comparing Ree-Hoover representations with the representations of Mayer 
coefficients of pressure and density expansions in series of degrees of activity
 The obtained results are presented in the tables and commented.
\end{abstract}

1. В статье рассматриваются термодинамические равновесные 
однокомпонентные системы клас\-си\-чес\-ких частиц, заключенных 
в $\nu$-мерном
дейст\-ви\-тель\-ном евклидовом пространстве ${\bf R}^{\nu}$ и
взаимодействующих посредством центральных сил, 
ха\-рак\-те\-ри\-зу\-е\-мых
потенциалом парного взаимодействия $\Phi({\bf r})$, где
${\bf r}=(r^{(1)}, r^{(2)}, \ldots,r^{(\nu)})\in {\bf R}^{\nu}$.
Предполагается, что потенциал парного взаимодействия $\Phi({\bf r})$
является измеримой функцией и удовлетворяет условиям устойчивости
[24, 17, 48] и регулярности [24, 17, 48].
Как обычно,  обозначим майеровскую функцию
\begin{equation}
f_{ij} = \exp\{-\beta\Phi({\bf r}_i-{\bf r}_j)\} - 1,
\end{equation}
где $i \ne j$, ${\bf r}_i,{\bf r}_j \in {\bf R}^{\nu}$,
$\beta=1/kT$ --- обратная температура, $k$ --- постоянная Больцмана,
\linebreak
$T$~---~аб\-со\-лют\-ная температура.
Через $\widetilde f_{ij}$ обозначим {\bf больцмановскую функцию}
[24, 48], полагая
\begin{equation}
\widetilde f_{ij} = 1 + f_{ij} = \exp\{-\beta\Phi({\bf r}_i-{\bf r}_j)\}.
\end{equation}
Зависимость дав\-ле\-ния~$p$ от плотности $\varrho$ в такой системе может
быть представлена в~двух видах: в форме вириального разложения давления
$p$
по степеням плотности $\varrho$ и в~параметрическом виде, т.е. в виде двух
уравнений, выражающих зависимость дав\-ле\-ния~$p$ и плотности $\varrho$
от параметра $z$, называемого активностью [22, 24, 43, 48].
Вириальное разложение имеет вид:
\begin{equation}
p = \beta^{-1}\sum_{n = 1}^{\infty}B_n(\beta)\varrho^n.
\end{equation}
Ниже мы будем для простоты опускать аргумент $\beta$ коэффициентов $B_n$.
В этом разложении коэффициенты
$B_n$ называются вириальными коэффициентами. Ви\-ри\-аль\-ный коэффициент $B_1$ равен
$1$, а при $n > 1$ вириальные коэффициенты опре\-де\-ля\-ют\-ся формулой:
\begin{equation}
B_n =-\frac{n-1}{n!}\sum_{B \in {\mathfrak B}_n}\int_{({\bf R}^{\nu})^{n-1}}
\prod_{\{u, v\} \in X(B)}f_{uv}(d{\bf r})_{1, n-1},
\end{equation}
где ${\mathfrak B}_n$ --- совокупность всех двусвязных $n$-вершинных
помеченных графов (блоков) с множеством вершин $V_n=\{1, 2, \ldots,n\}$,
$X(B)$ --- совокупность всех ребер блока $B$,
$(d{\bf r})_{1, n-1} = d{\bf r}_2d{\bf r}_3\ldots d{\bf r}_n$, \quad
$d{\bf r}_i=dr_i^{(1)}dr_i^{(2)}\cdot\ldots\cdot dr_i^{(\nu)}$.

Здесь и в дальнейшем изложении, следуя [25, 28], мы считаем, что
всякий граф $G$ по определению не имеет ни кратных ребер, ни петель.

Здесь и далее по тексту мы полагаем, что вершины ребер и графов
помечены натуральными числами. Поэтому всюду в статье мы отождествляем
вершины графов с их метками. Точно также мы отождествляем вершины,
инцидентные ребрам, с их метками.

Эти представления вириальных коэффициентов были получены Д. Майером
[22, 41--43].
Он же нашел и параметрическое представление зависимости дав\-ле\-ния~$p$
от плотности~$\varrho$:
\begin{equation}
p = \beta^{-1}\sum_{n = 1}^{\infty}b_n(\beta)z^n;
\end{equation}
\begin{equation}
\varrho = \sum_{n = 1}^{\infty}nb_n(\beta)z^n.
\end{equation}
Ниже мы будем для простоты опускать аргумент $\beta$ коэффициентов $b_n$.
В разложениях (5) и (6) коэффициенты
$b_n$ называются майеровскими коэффициентами (по степеням активности $z$).
Майеровский коэффициент $b_1$ равен
$1$, а при $n > 1$ майеровские коэффициенты $b_n$ определяются формулой:
\begin{equation}
b_n =\frac{1}{n!}\sum_{G \in {\bf G_n}}\int_{({\bf R}^{\nu})^{n-1}}
\prod_{\{u, v\} \in X(G)}f_{uv}(d{\bf r})_{1, n-1},
\end{equation}
где $\bf G_n$ --- совокупность всех связных $n$-вершинных
помеченных графов с множеством вершин
$V_n=\{1, 2, \ldots,n\}$, $X(G)$ --- совокупность всех ребер графа $G$.

Однако впоследствии было замечено, что эти представления  имеют
весьма не\-при\-ят\-ное свойство, благодаря которому они  практически непригодны
как для вычисления вириальных коэффициентов (за ис\-клю\-че\-ни\-ем первых трех),
так и для теоретического анализа
поведения старших коэффициентов.  Впервые на это свойство майеровских представлений
коэффициентов степенных рядов классической статистической ме\-ха\-ни\-ки
обратил внимание И.И. Иванчик. В своих работах
[1, 30] он впервые качественно описал это свойство и назвал его
{\bf асим\-п\-то\-ти\-че\-с\-кой катастрофой}. Она проявляется в~том, что
при майеровском представлении ко\-эф\-фи\-ци\-ен\-тов степенного ряда
зна\-чи\-тель\-ная часть слагаемых в сумме
интегралов,  определяющей $n$-ый коэффициент ряда,  с боль\-шой точностью
взаимно сокращаются
как величины противоположных знаков.

Сравнительно небольшой
оста\-ток, остающийся по\-с\-ле такого взаимного уни\-что\-же\-ния, является
при~$n\to\infty$ бесконечно малой величиной по~срав\-не\-нию
с числом слагаемых в сумме, традиционно определяющей этот
коэффициент. Этот "остаток", пред\-ста\-в\-ля\-ю\-щий основной интерес, даже
при небольших $n$ становится недоступным для не\-по\-сред\-с\-т\-вен\-но\-го
исследования.

В дальнейшем автор данной статьи в книге [17]
дал строгое математическое опре\-де\-ле\-ние асимптотической катастрофы.
Для удобства читателей приведем здесь это опре\-де\-ле\-ние.

О\,п\,р\,е\,д\,е\,л\,е\,н\,и\,е 1. В представлениях
коэффициентов степенного ряда
при\-сут\-ству\-ет феномен асимптотической катастрофы, если
при~любом $B>0$ число
слагаемых в~сумме,
представляющей коэффициент при
переменной в степени $n$,
при~$n\to\infty$ растет быстрее, чем величина
$(n!)^2B^n$. $\blacksquare$

Значение этого определения состоит в том, что оно дает возможность 
отделить те представления коэффициентов степенного ряда, где уже 
при сравнительно не\-боль\-ших $n$ число слагаемых слишком велико, 
от представлений, в которых число слагаемых растет значительно 
медленнее.

При попытках вычисления коэффициентов майеровских разложений, исходя из~та\-ких их
представлений, где присутствует феномен асимптотической катастрофы,
 прак\-ти\-че\-с\-ки неизбежно с ростом $n$ происходит
катастрофически быстрый рост ошибок вычислений этих коэффициентов.

В течение нескольких последних десятилетий усилия ряда ученых были 
направлены
на упрощение представлений коэффициентов степенных рядов классической
ста\-ти\-сти\-че\-с\-кой механики и их вычисление.

Целью упрощений представлений коэффициентов этих степенных рядов 
являлось упростить
процесс получения оценок этих коэффициентов с помощью их упрощенных 
представлений.
 Для краткости сложность процесса получения оценки данного
 коэффициента с помощью данного его представления мы будем называть 
{\bf сложностью данного представления этого коэффициента}.

Наиболее известными результатами в упрощении представлений вириальных
коэффициентов, по-видимому, являются представления Ри-Гувера [45],
[46], [47]. В этих представлениях при каждом $n \ge 4$ вириальный
коэффициент $B_n$ представляетcя линейной комбинацией интегралов,
маркированных полными помеченными графами, в которых каждое ребро помечено
либо майеровской, либо больцмановской функциями. В каждом интеграле,
являющемся членом такой линейной комбинации, подынтегральная функция
является произведением майеровских и больцмановских функций. А множество
всех майеровских и больцмановских функций, входящих в это произведение,
находится во взаимно однозначном соответствии с множеством ребер графа,
маркирующего этот интеграл. При этом каждому
помеченному майеровской функцией ребру этого графа соответствует
майеровская функция, входящая
множителем в это произведение. А каждому ребру, помеченному
больцмановской функцией, соответствует больцмановская функция, входящая
множителем в это произведение. Таким образом, вириальный
коэффициент $B_n$ представляетcя линейной комбинацией интегралов, в каждом
из которых подынтегральная функция является произведением майеровских и
больцмановских функций, общим числом ${n(n - 1)/2}$ функции. Эти
представления мы будем называть {\bf представлениями по методу Ри--Гувера}.

Используя представления Ри--Гувера вириальных
ко\-эф\-фи\-ци\-ен\-тов, рядом ученых  были вычислены [49] оценки 
значений вириальных коэффициентов $B_n$ (при $n=\overline{4,8})$ 
для ряда различных значений
температур. Позднее, на графическом компьютере были вычислены [50] для
по\-тен\-ци\-а\-ла Леннарда-Джонса  оценки значений вириальных коэффициентов $B_n$
при $n=\overline{6,9}$ для различных значений температур. В том числе были уточнены
ранее вычисленные оценки значений этих коэффициентов. Кроме того, при
$n=\overline{10,16}$ были вычислены оценки значений этих коэффициентов для нескольких
(от одного до четырех) значений температур. Кстати, то, что при $n=\overline{10,16}$
удалось найти оценку значения вириального коэффициента $B_n$ не более чем при
четырех различных значений температур, ука\-зы\-ва\-ет на то, что при $n>9$ объем
вычислений, необходимых для оценки одного из значений вириального
коэффициента $B_n$ по методу Ри-Гувера, так велик, что для этих вычислений требуется весьма
значительное время даже при работе на современном компьютере
с высокой про\-из\-во\-ди\-тель\-но\-с\-тью. Однако, остается открытым вопрос:
свободны ли представления Ри--Гувера от асим\-п\-то\-ти\-че\-с\-кой катастрофы.

Иной подход к упрощению представлений коэффициентов степенных рядов
клас\-си\-че\-с\-кой статистической механики разрабатывался автором данной
статьи. Он основан на раз\-ви\-ва\-е\-мой автором концепции каркасной
классификации помеченных графов \linebreak
\mbox{[2--8, 10, 13--20, 31--34, 37, 38]}.
Мы будем его называть {\bf методом каркасных сумм}. В рамках этого метода
им были получены свободные от асимптотической катастрофы представления
майеровских коэффициентов разложений давления и плотности по степеням
активности, представления коэффициентов раз\-ло\-же\-ния $m$-частичной
функции рас\-пре\-де\-ле\-ния в ряд по степеням
активности, представления коэффициентов разложения отношения активности
к плотности в ряд по степеням активности и представления вириальных
коэффициентов [3, 4, 6--8, 10, 15, 17, 31--34, 37].

Достоинством этих представлений является то, что они свободны от
асим\-п\-то\-ти\-че\-с\-кой катастрофы 
[10, 11, 15, 17, 36, 37]. Используя эти
представления, удалось получить [9, 10, 12, 17, 35] оценку сверху радиуса 
сходимости майеровских разложений по степеням активности 
(для не\-от\-ри\-ца\-тель\-но\-го потенциала).
А также удалось, используя эти представления, на бытовом компьютере
вы\-чи\-с\-лить, довольно точно, оценки 4-ого, 5-ого и 6-ого
вириальных коэффициентов при одном из значений температуры.

2. Целью статьи является: определить критерии для оценки сложности
представлений ко\-эф\-фи\-ци\-ен\-тов степенных рядов классической
статистической механики;
про\-де\-мон\-стри\-ро\-вать их  применение на примерах сравнения
представлений Ри-Гувера вириальных коэффициентов и представлений
ко\-эф\-фи\-ци\-ен\-тов степенных рядов, основанных на концепции
каркасной классификации помеченных графов.

Очевидно, что даже для сравнения по сложности двух различных
представлений данного ко\-эф\-фи\-ци\-ен\-та некоторого степенного ряда
надо иметь критерий. Тем более необходим такого рода критерий, если
поставлена задача сравнить по сложности данные представления
ко\-эф\-фи\-ци\-ен\-тов при переменной в степени $n$ двух различных
степенных рядов.
Создание такого рода критериев облегчает то обстоятельство,
что многие известные представления ко\-эф\-фи\-ци\-ен\-тов степенных рядов
классической статистической механики являются линейными комбинациями
многомерных интегралов, маркированных помеченными графами, в которых
каждое ребро помечено либо майеровской, либо больцмановской функциями.
В каждом интеграле, являющемся членом такой линейной
комбинации, подынтегральная функция является произведением майеровских и
больцмановских функций (таковы, например, предложенные Ри и Гувером
[45, 46, 47] представления вириальных коэффициентов).

В этой статье произведена классификация представлений коэффициентов
степенных рядов классической ста\-ти\-сти\-че\-с\-кой механики. Наиболее
важный класс этой классификации содержит все те упомянутые выше
представления коэффициентов степенных рядов классической
ста\-ти\-сти\-че\-с\-кой механики, которые являются линейными комбинациями
многомерных интегралов, описанными в предыдущем абзаце.

Для оценки сравнительной сложности входящих в этот класс представлений
коэффициентов степенных рядов в этой статье построены три критерия,
упорядоченные по их точности.
При\-ме\-не\-ние этих критериев продемонстрировано на примерах оценки
сравнительной сложности представлений Ри-Гувера вириальных коэффициентов
и представлений ко\-эф\-фи\-ци\-ен\-тов степенных рядов, основанных
на концепции каркасной классификации помеченных графов.

3. Прежде, чем перейти к описанию предлагаемой классификации и 
предлагаемых
критериев сложности представлений ко\-эф\-фи\-ци\-ен\-тов степенных 
рядов, дадим определения математических понятий, необходимых для их 
описания, и остановимся на некоторых свойствах этих понятий.

Прежде всего мы несколько расширим понятие ребра помеченного графа, 
введя следующее

О\,п\,р\,е\,д\,е\,л\,е\,н\,и\,е 2. Неупорядоченная пара $\{i,j\}$
различных натуральных чисел называется {\bf ребром}. $\blacksquare$

В данной статье мы будем рассматривать только множества попарно 
различных ребер, не оговаривая это обстоятельство.

О\,п\,р\,е\,д\,е\,л\,е\,н\,и\,е 3. Будем говорить, что множество
ребер $X_f = \{\{i,j\}\}$ {\bf определяет} множество
$F = \{f_{i j}\}$ майеровских функций, если майеровская функция
$f_{i j}$ входит во множество $F$ тогда и только тогда, когда
ребро $\{i,j\}$ принадлежит  множеству $X_f$. При этом множество
ребер $X_f$ будем называть {\bf множеством майеровских ребер по отношению
к этому множеству $F$ майеровских функций}. $\blacksquare$

О\,п\,р\,е\,д\,е\,л\,е\,н\,и\,е 4. Будем также говорить, что множество
ребер $X_{\widetilde f} = \{\{i',j'\}\}$ {\bf определяет} множество
\mbox{$\widetilde F = \{\widetilde f_{i' j'}\}$} больцмановских
функций, если больцмановская функция $\widetilde f_{i' j'} = f_{i' j'} + 1$
содержится во множество $\widetilde F$ тогда и только тогда, когда ребро
$\{i',j'\}$ принадлежит множеству $X_{\widetilde f}$. При этом
множество  $X_{\widetilde f}$ будем называть {\bf множеством больцмановских
ребер по отношению к этому множеству $\widetilde F$ больцмановских 
функций}. $\blacksquare$

Введем обозначения:
\begin{equation}
P(F, \widetilde F) = \prod_{f_{i j}\in F}
\prod_{\widetilde f_{i' j'}\in \widetilde F} f_{i j}\widetilde f_{i' j'}
\label{p}
\end{equation}
--- произведение всех майеровских функций, принадлежащих множеству
майеровских функций $F$, и всех больцмановских функций, принадлежащих
множеству больцмановских функций $\widetilde F$. Очевидно, что произведение
$P(F, \widetilde F)$ является функцией множеств $F$ и $\widetilde F$. Для
краткости письма мы будем опускать аргументы $F$ и $\widetilde F$
произведения $P$. А произведение $P$ будем называть {\bf произведением
майеровских и больцмановских функций}.

${\bf X} = \{X_f, X_{\widetilde f}\}$ --- упорядоченная пара
непересекающихся множеств: множества ребер $X_f = \{\{i,j\}\}$ и множества
ребер $X_{\widetilde f} = \{\{i',j'\}\}$.

$V(X_f)$ --- множество концов (вершин) всех ребер из множества $X_f$.

$V(X_{\widetilde f})$ --- множество концов (вершин) всех ребер из множества
$X_{\widetilde f}$.

$\left|V(X_f)\bigcup V(X_{\widetilde f})\right|$ --- мощность суммы
множеств $V(X_f)$ и $V(X_{\widetilde f})$.

Ниже мы будем рассматривать и такие упорядоченные пары
${\bf X} = \{X_f, X_{\widetilde f}\}$ непересекающихся множеств,
в которых второе множество является пустым множеством, то есть
пары, имеющие вид ${\bf X} = \{X_f, \emptyset\}$.

О\,п\,р\,е\,д\,е\,л\,е\,н\,и\,е 5. Если множества ребер $X_f$ и
$X_{\widetilde f}$ удовлетворяют условию
\begin{equation}
V(X_f)\bigcup V(X_{\widetilde f}) = V_n = \{1, 2, \ldots, n\},
\end{equation}
где
\begin{equation}
n = \left|V(X_f)\bigcup V(X_{\widetilde f})\right|,
\end{equation}
то упорядоченную пару ${\bf X}=\{X_f, X_{\widetilde f}\}$ этих множеств
будем называть {\bf канонической парой множеств}, а число $n$ ---
{\bf порядком} этой канонической пары множеств.
В канонической паре множеств ${\bf X}=\{X_f, X_{\widetilde f}\}$ первое
множество $X_f$ будем называть {\bf множеством майеровских ребер},
а второе множество $X_{\widetilde f}$ --- {\bf множеством больцмановских
ребер}. $\blacksquare$

Через ${\mathfrak X}_n = \{{\bf X}=(X_f,\, X_{\widetilde f})\}$ обозначим
совокупность всех канонических пар множеств порядка $n$.
Заметим, что в паре ${\bf X}=(X_f, X_{\widetilde f})$, входящей
в совокупность ${\mathfrak X}_n$, множество больцмановских ребер
$X_{\widetilde f}$ может быть и пустым.

Каждой канонической паре множеств ${\bf X}=(X_f, X_{\widetilde f})$
порядка $n$ поставим в соответствие произведение майеровских и
больцмановских функций $P_n({\bf X})$, определенное формулой
\begin{equation}
 P_n({\bf X}) = \prod_{\{i,j\} \in X_f({\bf X})}
\prod_{\{i',j'\} \in X_{\widetilde f}({\bf X})}f_{ij}\widetilde f_{i'j'}.
\label{8}
\end{equation}

Очевидно, что произведение майеровских и больцмановских функций $P_n({\bf X})$ 
является сужением на множество ${\mathfrak X}_n$ функции $P(F, \widetilde F)$, 
определенной формулой (8).

О\,п\,р\,е\,д\,е\,л\,е\,н\,и\,е 6. Будем говорить, что каноническая пара
множеств ${\bf X}= (X_f, X_{\widetilde f})$ порядка $n$ {\bf определяет}
произведение функций $P_n({\bf X})$ и называть это произведение
функций {\bf каноническим произведением}, а
число $n$ --- {\bf порядком} этого произведения. $\blacksquare$

Через
${\mathfrak P}_n = \{P \colon P =
P_n({\bf X}), \; {\bf X}\in {\mathfrak X}_n \}$
обозначим множество всех канонических произведений, определенных
каноническими парами множеств из совокупности ${\mathfrak X}_n$.

Из определения совокупности ${\mathfrak X}_n$, множества ${\mathfrak P}_n$
и произведения $P_n({\bf X})$ формулой (11) следует, что соотношение
\begin{equation}
P = P_n({\bf X})
\label{12}
\end{equation}
между элементами ${\bf X} \in {\mathfrak X}_n$ и
$P \in {\mathfrak P}_n$ является отображением совокупности
${\mathfrak X}_n = \{{\bf X}\}$ на множество ${\mathfrak P}_n = \{P\}$.

Заметим, что так определенное отображение
$P_n\colon {\mathfrak X}_n \to {\mathfrak P}_n$ является
взаимно однозначным отображением совокупности ${\mathfrak X}_n$
на множество ${\mathfrak P}_n$.
Так как каждое  произведение функций $P$ из множества
${\mathfrak P}_n$ при отображении $P_n$ имеет, и при том единственный,
прообраз
${\bf X} = (X_f, X_{\widetilde f})$ в совокупности ${\mathfrak X}_n$, то
этот прообраз можно принять за метку этого произведения и считать это
произведение помеченным канонической парой множеств
${\bf X} = (X_f, X_{\widetilde f})$. При этом всякая каноническая пара
множеств ${\bf X} = (X_f, X_{\widetilde f})$ из совокупности
${\mathfrak X}_n$ окажется меткой канонического произведения функций,
которое входит во множество ${\mathfrak P}_n$ и однозначно
определяется этой парой множеств по формулам (12) и (11). Ниже
будут изложены
и другие способы пометки канонических произведений функций, нашедшие свое
приложение в данной статье.

Обозначим через
${\mathfrak G}_n = \{G(V_n;\, X_f,\, X_{\widetilde f})\}$ множество
помеченных графов с множеством вершин $V_n = \{1, 2, \ldots,n\}$ и
множеством ребер $X$, являющимся  объединением двух непересекающихся
множеств: множества $X_f = \{\{i,j\}\}$ и множества
$X_{\widetilde f} = \{\{i',j'\}\}$, --- образующих каноническую пару
множеств $(X_f, X_{\widetilde f}) \in {\mathfrak X}_n$.

Для графов, принадлежащих множеству
${\mathfrak G}_n = \{G(V_n;\, X_f,\, X_{\widetilde f})\}$,
введем обо\-зна\-че\-ния:

$X_f(G)$ --- множество майеровских ребер графа $G\in {\mathfrak G}_n$;

$X_{\widetilde f}(G)$ --- множество больцмановских ребер графа
$G\in {\mathfrak G}_n$.

Определим отображение $A_n$ множества ${\mathfrak G}_n$ на множество
 ${\mathfrak X}_n$, полагая
\begin{equation}
A_n(G) = (X_f(G), X_{\widetilde f}(G)).
\label{13}
\end{equation}
Отображение $A_n$, определенное формулой (13) является взаимно
однозначным отображением множества ${\mathfrak G}_n$ на множество
${\mathfrak X}_n$.

Напомним, что отображение $P_n$, определенное формулами (11)
и (12),
является отображением множества ${\mathfrak X}_n$
на множество ${\mathfrak P}_n$. Значит, существует композиция отображений
$P_n\circ A_n$, являющееся отображением множества
 ${\mathfrak G}_n$ на множество ${\mathfrak P}_n$.

Так как отображения $A_n$ и $P_n$ --- взаимно однозначные отображения, то и
их композиция $P_n\circ A_n$ также является [23, 39] взаимно
однозначным отображением.

{\bf Замечание 1.} Каждое  произведение функций $P$ из множества 
${\mathfrak P}_n$ при отображении $P_n\circ A_n$ имеет, и при том 
единственный, прообраз
во множестве ${\mathfrak G}_n$. Значит, этот прообраз можно принять
за граф-метку этого произведения и считать это произведение помеченным.
При этом всякий граф $G(V_n; X_f, X_{\widetilde f})$
из множества ${\mathfrak G}_n$ окажется меткой произведения функций,
которое мы обозначим $P_{1n}(G)$. Это произведение входит
во множество ${\mathfrak P}_n$ и однозначно определяется этим графом
по формуле
\begin{multline}
P_{1n}(G) = (P_n\circ A_n)(G) = P_n(A_n(G)) =
P_n((X_f(G), X_{\widetilde f}(G))) = \\
\prod_{\{i,j\} \in X_f(G)}\prod_{\{i',j'\} \in X_{\widetilde f}(G)}
f_{ij}\widetilde f_{i'j'}.
\label{14}  
\end{multline} 
\hbox to \textwidth{\hfil \raisebox{5pt}[0pt][0pt]{$\blacksquare$}}

Опираясь на замечание 1, сформулируем следующее

О\,п\,р\,е\,д\,е\,л\,е\,н\,и\,е 7. Если граф
$G(V_n;\, X_f,\, X_{\widetilde f})$
принадлежит множеству ${\mathfrak G}_n$, то каноническое произведение
функций $P_{1n}(G)$, определенное формулой (14) будем называть
{\bf произведением, помеченным графом}
$G = G(V_n; X_f, X_{\widetilde f})$, а граф
$G = G(V_n; X_f, X_{\widetilde f})$ --- {\bf графом-меткой} этого
произведения функций. $\blacksquare$

Обозначим $R(G) = (V_n; X_f)$ граф с множеством вершин $V_n$ и множеством
ребер
$X_f$, являющийся подграфом графа $G = G(V_n; X_f, X_{\widetilde f})$,
принадлежащего множеству графов ${\mathfrak G}_n$. Множеством
ребер графа $R(G)$ по определению является множество $X_f(G)$ майеровских
ребер графа $G$. Это множество ребер определяет множество майеровских
функций, входящих в произведение функций $P_{1n}(G)$.
Но граф $R(G)$ по определению не содержит, в отличие от графа $G$,
множества $X_{\widetilde f}(G)$ больцмановских ребер, определяющего
множество больцмановских функций, входящих в произведение функций
$P_{1n}(G)$. Поэтому будем называть
подграф $R(G)$ графа $G$ {\bf недостаточной меткой} произведения функций
$P_{1n}(G)$, помеченного графом $G$.

О\,п\,р\,е\,д\,е\,л\,е\,н\,и\,е 8. Произведение функций 
$P \in {\mathfrak P}_n$ будем называть {\bf базовым произведением порядка $n$},
если его граф-метка $G \in {\mathfrak G}_n$ удовлетворяет условию:
подграф $R(G)$ графа $G$ является связным графом. Если же под\-граф
$R(G)$ графа $G$ не является
связным, то произведение функций $P$\, будем называть {\bf псевдобазовым
произведением}. $\blacksquare$

{\bf Лемма 1.} {\it Если подграф $R(G)$ графа-метки
$G \in {\mathfrak G}_n$
является связным, то, во-первых, каждое ребро из множества $X_{\widetilde f}(G)$
соединяет две несмежные  вершины графа $R(G)$ и, во-вторых, помеченное
графом $G$
каноническое произведение $P_{1n}(G)$ является функцией $n$ переменных
${\bf r}_1, {\bf r}_2, \ldots, {\bf r}_n$.}

{\bf Доказательство.} Так как любое ребро из множества
$X_{\widetilde f}(G)$ принадлежит графу $G$ по определению этого графа, то
обе инцидентные ему вершины принадлежат множеству $V_n$.
Следовательно, эти вершины принадлежат графу $R(G)$ по его определению.
По условиям леммы граф $G$ принадлежит множеству ${\mathfrak G}_n$. Отсюда
по определению этого множества следует, что множества $X_{\widetilde f}$
и $X_f$ не имеют общих ребер и образуют каноническую пару порядка $n$.
Это означает, что множество $X_f$ не содержит
ребра, соединяющего  две вершины, инцидентные какому-нибудь ребру из
множества $X_{\widetilde f}(G)$. Значит, каждое ребро из множества
$X_{\widetilde f}$ соединяет две несмежные вершины графа $R(G)$. Первое
утверждение леммы доказано.

Докажем теперь второе утверждение леммы.
Пусть $i$ --- вершина, принадлежащая множеству $V_n$. Так как
подграф $R(G) = (V_n; X_f)$ графа $G$ является связным, то во множестве
ребер $X_f(G)$ существует ребро, соединяющее вершину $i$ с некоей
вершиной $j \in V_n$. Отсюда по определению произведения $P_{1n}(G)$
формулой (14) следует, что майеровская функция $f_{i j}$ входит
в это произведение. А так как майеровская  функция $f_{i j}$ является
по определению функцией переменных ${\bf r}_i$ и ${\bf r}_j$,
то эти переменные входят в число переменных произведения
функций $P_{1n}(G)$. Таким образом, при любом $i \in V_n$
переменная ${\bf r}_i$ входят в число переменных функции, являющейся
произведением функций $P_{1n}(G)$.

С другой стороны, если $i \notin V_n$, то $i$ не является вершиной графа
$G$ и не может быть вершиной, инцидентной какому-либо ребру этого графа.
Поэтому из определения произведения $P_{1n}(G)$ следует, что
переменная ${\bf r}_i$ не является переменной какой-либо из функций,
входящих в это произведение. Из полученных результатов следует второе
утверждение леммы. $\blacktriangleright$

Из леммы 1 вытекает следующее

{\bf Следствие 1.} {\it Базовое произведение $P \in {\mathfrak P}_n$
является функцией $n$ переменных
${\bf r}_1, {\bf r}_2, \ldots, {\bf r}_n$,
где $n$ --- число вершин графа-метки $G$.} 

О\,п\,р\,е\,д\,е\,л\,е\,н\,и\,е 9. Если произведение функций 
$P \in {\mathfrak P}_n$ является базовым порядка $n$, то интеграл
\begin{equation}
I(P) = \int_{({\bf R}^{\nu})^{n-1}} P(d{\bf r})_{1, n-1}
\label{15}
\end{equation}
будем называть {\bf базовым интегралом}, а число $n$ --- его
{\bf порядком}. $\blacksquare$

{\bf Замечание 2.} Базовый интеграл $I(P)$ от базового произведения
$P \in {\mathfrak P}_n$ полностью определяется этим произведением. 
Поэтому можно считать, что граф-метка $G$ этого базового произведения 
является также и графом-меткой базового интеграла $I(P)$. 
$\blacksquare$

Базовый интеграл, помеченный графом-меткой $G \in {\mathfrak G}_n$, 
обозначим $I(G)$, полагая
\begin{equation}
I(G) = I(P_{1n}(G)) = 
\int_{({\bf R}^{\nu})^{n-1}} P_{1n}(G)(d{\bf r})_{1, n-1}
\label{15'}
\end{equation}

О\,п\,р\,е\,д\,е\,л\,е\,н\,и\,е 10. Интеграл от псевдобазового произведения функций
будем называть {\bf псевдобазовым интегралом}. $\blacksquare$

О\,п\,р\,е\,д\,е\,л\,е\,н\,и\,е 11. Линейная комбинация базовых интегралов
порядка $n$, в которой коэффициент при каждом из входящих в нее
интегралов является
действительным числом и определяется маркирующим этот интеграл графом,
называется {\bf базовой линейной комбинацией}, а число $n$ называется ее
{\bf порядком}. $\blacksquare$

О\,п\,р\,е\,д\,е\,л\,е\,н\,и\,е 12. Линейная комбинация интегралов
от произведений майеровских и больцмановских функций, в которой хотя бы
один интеграл не является базовым, называется {\bf псевдобазовой линейной
комбинацией}. $\blacksquare$

{\bf Пример 1}
По определению 9 при любом натуральном $n \ge 2$
представление Ри-Гувера [47] вириального коэффициента $B_n$ является
линейной комбинацией базовых интегралов порядка $n$. В этой линейной
комбинации коэффициент при каждом из входящих в нее интегралов является
действительным числом и определяется маркирующим этот интеграл графом.
Отсюда по определению 11 следует, что при любом натуральном $n \ge 2$
представление Ри-Гувера вириального коэффициента $B_n$ является базовой
линейной комбинацией порядка $n$. $\blacktriangleright$

О\,п\,р\,е\,д\,е\,л\,е\,н\,и\,е 13. Число интегралов, входящих в базовую
линейную комбинацию $L$, называется {\bf длиной} этой линейной комбинации и
обозначается $q$. $\blacksquare$

Введем обозначения:

${\mathfrak G}(L)$ --- множество всех графов, служащих графами-метками
таких базовых произведений, которые являются подынтегральными функциями
интегралов, входящих в базовую линейную комбинацию $L$;

\begin{equation}
R({\mathfrak G}(L)) = \{R(G): \quad G \in {\mathfrak G}(L)\}.
\end{equation}

Часто встречаются случаи, когда для маркировки канонического произведения
функций $P \in {\mathfrak P}_n$ проще использовать не граф-метку такого
произведения функций, а другие графы. Например, граф
$\widetilde G = \widetilde G(V_n, X_f)$, где $X_f$  --- множество
майеровских ребер по отношению к множеству $F$ всех майеровских функций,
входящих в данное каноническое произведение функций
$P \in {\mathfrak P}_n$.

Граф $\widetilde G(V_n, X_f)$ дает возможность непосредственно
опре\-де\-лить только майеровские функции, входящие в произведение
функций $P(X_f, X_{\widetilde f})$. Для определения же больцмановских
функций, входящих в это произведение, в ряде случаев
пред\-по\-чти\-тель\-нее, в обход определения графа-метки этого
произведения, пря\-мо указать множество $X_{\widetilde f}$\,
больцмановских ребер по отношению к множеству $\widetilde F$ всех
больцмановских функций,
входящих в данное каноническое произведение функций
$P \in {\mathfrak P}_n$, или же указать
конструктивный метод построения этого множества. Это дает
возможность непосредственно определить больцмановские функции, входящие
в помеченное графом $\widetilde G$ произведение функций.
Множество $X_{\widetilde f}$ дополняет множество ребер графа
$\widetilde G$
до множества ребер графа-метки этого произведения. Назовем это множество
{\bf дополняющим} и обозначим $X_{\rm ad}(\widetilde G)$, полагая
$X_{\rm ad}(\widetilde G) = X_{\widetilde f}$.

Обозначим $\widetilde{\mathfrak G}_n = \{\widetilde G\}$, где $n \ge 3$,
конечное множество попарно различных связных помеченных графов
с множеством вершин $V_n$, каждому из которых по\-став\-ле\-но
в соответствие дополняющее множество $X_{\rm ad}(\widetilde G)$, которое
не пересекается с множеством
$X_f(\widetilde G)$ майеровских ребер и образует c ним  каноническую пару
$(X_f(\widetilde G), X_{\rm ad}(\widetilde G))\in {\mathfrak X}_n$.

О\,п\,р\,е\,д\,е\,л\,е\,н\,и\,е 14. Графы из множества
$\widetilde{\mathfrak G}_n$ будем называть
{\bf уком\-плек\-то\-ван\-ны\-ми}. $\blacksquare$

Введем обозначения:

$\widetilde {\mathfrak X}_n =
\{(X_f(\widetilde G), X_{\rm ad}(\widetilde G)) \colon
\widetilde G \in \widetilde{\mathfrak G}_n\}$

$\widetilde {\mathfrak P}_n =
P_n(\widetilde{\mathfrak X}_n)$ \. --- \,
образ множества канонических пар $\widetilde{\mathfrak X}_n \subset
{\mathfrak X}_n$ при отображении
\mbox{$P_n\colon {\mathfrak X}_n \to {\mathfrak P}_n$};

${\widetilde P_n} = P_n \mid_{\widetilde{\mathfrak X}_n}$ --- сужение
отображения $P_n$ на подмножество $\widetilde {\mathfrak X}_n \subset
{\mathfrak X}_n$.

По определению отображение ${\widetilde P}_n$ является взаимно однозначным
отображением множества $\widetilde{\mathfrak X}_n$ на множество
$\widetilde{\mathfrak P}_n$.

Определим отображение $\widetilde A_n$ множества
$\widetilde{\mathfrak G}_n$ на множество $\widetilde{\mathfrak X}_n$,
полагая
\begin{equation}
{\widetilde A}_n(\widetilde G) =
(X_f(\widetilde G), X_{\rm ad}(\widetilde G)), \quad
\widetilde G \in \widetilde{\mathfrak G}_n.
\label{13'}
\end{equation}
Отображение ${\widetilde A}_n$, определенное формулой (18) является
взаимно однозначным отображением множества $\widetilde{\mathfrak G}_n$
на множество $\widetilde{\mathfrak X}_n$.

{\bf Замечание 3.}  Так как область определения отображения 
${\widetilde P}_n$ совпадает с областью значений отображения 
${\widetilde A}_n$, то композиция отображений 
${\widetilde P}_n\circ {\widetilde A}_n$ существует и является  
отображением множества  $\widetilde{\mathfrak G}_n$ на множество 
$\widetilde{\mathfrak P}_n$. 

Так как отображения ${\widetilde A}_n\colon
\widetilde{\mathfrak G}_n \to \widetilde{\mathfrak X}_n$ и 
${\widetilde P}_n\colon
\widetilde{\mathfrak X}_n \to \widetilde{\mathfrak P}_n$ --- взаимно
однозначные, то их композиция 
${\widetilde P}_n\circ {\widetilde A}_n\colon
\widetilde{\mathfrak G}_n \to \widetilde{\mathfrak P}_n$
является [23, 39] взаимно однозначным отображением множества  
$\widetilde{\mathfrak G}_n$ на множество $\widetilde{\mathfrak P}_n$.
$\blacksquare$

Из замечания 3 вытекает 

{\bf Следствие 2.} {\it Каждое  произведение функций $\widetilde P$ 
из множества $\widetilde{\mathfrak P}_n$
при отображении ${\widetilde P}_n\circ {\widetilde A}_n$ имеет, 
и при том единственный, прообраз
во множестве $\widetilde{\mathfrak G}_n$. Значит, этот прообраз 
является графом, который можно
принять за метку этого произведения и считать это произведение
помеченным этим графом. При этом всякий граф $\widetilde G$
из множества $\widetilde{\mathfrak G}_n$ окажется меткой произведения
функций, которое является образом этого графа при отображении
${\widetilde P}_n\circ {\widetilde A}_n\colon
\widetilde{\mathfrak G}_n \to \widetilde{\mathfrak P}_n$.} 

Образ графа $\widetilde{G} \in \widetilde{\mathfrak G}_n$  при отображении
${\widetilde P}_n\circ {\widetilde A}_n\colon
\widetilde{\mathfrak G}_n \to \widetilde{\mathfrak P}_n$ обозначим 
$\widetilde P_{1n}(\widetilde G)$.

Опираясь на замечание 3 и следствие 2, сформулируем следующее

О\,п\,р\,е\,д\,е\,л\,е\,н\,и\,е 15. Произведение функций 
$\widetilde{P}_{1n}(\widetilde G)$, являющееся образом графа 
  $\widetilde{G}(V_n, X_f) \in \widetilde{\mathfrak G}_n$ при отображении
\mbox{${\widetilde P}_n\circ {\widetilde A}_n\colon
\widetilde{\mathfrak G}_n \to \widetilde{\mathfrak P}_n$}, будем называть
{\bf произведением, по\-ме\-чен\-ным графом}
$\widetilde G = \widetilde G(V_n, X_f)$, а граф
$\widetilde{G}(V_n, X_f)$ --- {\bf укомплектованным гра\-фом-мет\-кой} этого
произведения. $\blacksquare$

{\bf Лемма 2.} Если граф $\widetilde{G}(V_n, X_f)$
принадлежит множеству $\widetilde{\mathfrak G}_n$, то помеченное этим графом 
произведение функций $\widetilde{P}_{1n}(\widetilde G)$ является каноническим 
произведением порядка $n$ и представляется формулой
\begin{equation}
\widetilde P_{1n}(\widetilde G) =
\prod_{\{i,j\} \in X_f(G)}\prod_{\{i',j'\} \in X_{\rm ad}(\widetilde G)}
f_{ij}\widetilde f_{i'j'}.
\label{14'}
\end{equation}

{\bf Доказательство.} Докажем сначала, что произведение функций 
$\widetilde{P}_{1n}(\widetilde G)$ является каноническим произведением порядка $n$.
Из определения множества $\widetilde{\mathfrak P}_n$ следует, что это множество
является подмножеством множества ${\mathfrak P}_n$ канонических произведений 
порядка $n$. Отсюда и из замечания 3 следует, множество $\widetilde{\mathfrak P}_n$ 
значений отображения ${\widetilde P}_n\circ {\widetilde A}_n\colon
\widetilde{\mathfrak G}_n \to \widetilde{\mathfrak P}_n$ является множеством 
 канонических произведений порядка $n$. Стало быть, каков бы ни был граф 
$\widetilde{G}(V_n, X_f) \in \widetilde{\mathfrak G}_n$, его образ 
$\widetilde P_{1n}(\widetilde G)$ при отображении 
${\widetilde P}_n\circ {\widetilde A}_n\colon
\widetilde{\mathfrak G}_n \to \widetilde{\mathfrak P}_n$ является каноническим 
произведением порядка $n$. По определению 15 произведение 
$\widetilde P_{1n}(\widetilde G)$ называется произведением, помеченным графом
$\widetilde G$. Итак, доказано, что помеченное графом 
$\widetilde{G} \in \widetilde{\mathfrak G}_n$ произведение функций 
$\widetilde{P}_{1n}(\widetilde G)$ является каноническим произведением порядка $n$.

Докажем теперь, что помеченное графом $\widetilde{G} \in \widetilde{\mathfrak G}_n$
произведение функций $\widetilde{P}_{1n}(\widetilde G)$ представляется формулой (19).
Из определения произведения функций 
$\widetilde P_{1n}(\widetilde G)$, определения отображения ${\widetilde P}_n\colon
\widetilde{\mathfrak X}_n \to \widetilde{\mathfrak P}_n$, определения отображения 
$P_n\colon {\mathfrak X}_n \to {\mathfrak P}_n$ формулами (11) и (12) и 
определения отображения ${\widetilde A}_n\colon
\widetilde{\mathfrak G}_n \to \widetilde{\mathfrak X}_n$ формулой (18)
 следует
\begin{multline}
\widetilde P_{1n}(\widetilde G) =
\widetilde P_n\circ \widetilde A_n(\widetilde G) =
\widetilde P_n(\widetilde A_n(\widetilde G)) =
\widetilde P_n((X_f(\widetilde G), X_{\rm ad}(\widetilde G) = \\
P_n((X_f(G), X_{\rm ad}(\widetilde G))) =
\prod_{\{i,j\} \in X_f(G)}\prod_{\{i',j'\} \in X_{\rm ad}(\widetilde G)}
f_{ij}\widetilde f_{i'j'}.
\label{14'.1}
\end{multline}
Отсюда следует формула (19). Лемма 2 полностью доказана. 
$\blacktriangleright$

{\bf Теорема 1.} {\it Если граф $\widetilde G(V_n, X_f)$, которому
поставлено в соответствие дополняющее множество
$X_{\rm ad}(\widetilde G)$, принадлежит множеству
$\widetilde{\mathfrak G}_n$, то верны следующие утверждения:

$A_1$. Граф $G(V_n;\, X_f(\widetilde G),\, X_{\rm ad}(\widetilde G))$,
принадлежит множеству ${\mathfrak G}_n$ и является меткой произведения
${\widetilde P}_{1n}(\widetilde G)$.

$A_2$. Граф $\widetilde G$  является образом графа-метки
$G(V_n;\, X_f(\widetilde G),\, X_{\rm ad}(\widetilde G))$
при отображении $R$.

$A_3$. Произведение $\widetilde P_{1n}(\widetilde G)$ майеровских и
больцмановских функций является базовым произведением порядка $n$,
а граф $\widetilde G$ --- его
укомплектованным графом-меткой.}

{\bf Доказательство.} По определению множества
$\widetilde{\mathfrak G}_n$ дополняющее
множество $X_{\rm ad}(\widetilde G)$ образует с множеством ребер
$X_f(\widetilde G)$ каноническую пару
$(X_f(\widetilde G), X_{\rm ad}(\widetilde G))\in {\mathfrak X}_n$ .

Отсюда следует, что граф
$G(V_n;\, X_f(\widetilde G),\, X_{\rm ad}(\widetilde G))$
 принадлежит множеству графов ${\mathfrak G}_n$ по определению этого
множества. По определению 7, помеченное этим графом каноническое
произведение функций $P_{1n}(G)$, определяется формулой (14),
которая в данном случае имеет вид
\begin{multline}
P_{1n}(G) = (P_n\circ A_n)(G) = P_n(A_n(G)) =
P_n((X_f(\widetilde G), X_{\rm ad}(\widetilde G))) = \\
\prod_{\{i,j\} \in X_f(G)}\prod_{\{i',j'\} \in X_{\rm ad}(G)}
f_{ij}\widetilde f_{i'j'}.
\label{14"}
\end{multline}
А каноническое произведение функций $\widetilde{P}_{1n}(\widetilde G)$
определяется формулой (19).
Из формул (21) и (19) следует равенство
$P_{1n}(G) = \widetilde P_{1n}(\widetilde G)$. Отсюда по определению 7
следует, что граф
$G(V_n;\, X_f(\widetilde G),\, X_{\rm ad}(\widetilde G))$, является
графом-меткой произведения ${\widetilde P}_{1n}(\widetilde G)$.
Утверждение $A_1$ полностью доказано.

Из определения графа $\widetilde G$, утверждения $A_1$ и
определения отображения $R$ следует утверждение $A_2$.

Так как граф $\widetilde G$ --- связный, то из утверждений $A_1$ и $A_2$
по определению 8 следует, что произведение
$\widetilde P_{1n}(\widetilde G)$ является базовым порядка $n$.
По определению 15 из условий теоремы 1 следует, что граф $\widetilde G$
является укомплектованным графом-меткой этого произведения. Утверждение
$A_3$ доказано. Теорема 1 полностью доказана. $\blacktriangleright$

Для каждого графа ${\widetilde G} \in \widetilde{\mathfrak G}_n$ определим
интеграл $\widetilde I(\widetilde G)$, полагая
\begin{equation}
\widetilde I(\widetilde G) = \int_{({\bf R}^{\nu})^{n-1}}
{\widetilde P}_{1n}(\widetilde G)(d{\bf r})_{1, n-1}.
\label{22} 
\end{equation} 

{\bf Замечание 4.} По теореме 1 произведение функций
$\widetilde P_{1n}(\widetilde G)$, определенное формулой (19),
является базовым порядка $n$, а граф $\widetilde G$ является его укомплектованным
графом-меткой. Базовый интеграл $\widetilde I(\widetilde G)$ порядка $n$
от базового произведения $\widetilde P_{1n}(\widetilde G)$ порядка $n$ полностью
определяется этим произведением. А так как это произведение помечено
графом $\widetilde G$, то интеграл $\widetilde I(\widetilde G)$ полностью
определяется этим графом. Поэтому можно считать, что граф $\widetilde G$
является укомплектованным графом-меткой не только базового произведения
$\widetilde P_{1n}(\widetilde G)$ порядка $n$, но и базового интеграла
$\widetilde I(\widetilde G)$ порядка $n$. $\blacksquare$

{\bf Теорема 2.} {\it Пусть непустое подмножество графов
$\widetilde{\mathfrak G}_n^{(0)}$ множества графов
$\widetilde{\mathfrak G}_n$ удовлетворяет условию:
для каждого графа
$\widetilde G(V_n; X_f) \in \widetilde{\mathfrak G}_n^{(0)}$
определен поставленный ему в соответствие коэффициент
$c(\widetilde G)$, являющийся
действительным числом.

Тогда линейная комбинация интегралов
\begin{equation}
L = \sum_{\widetilde G\in\widetilde{\mathfrak G}_n^{(0)}}
c(\widetilde G)\widetilde I(\widetilde G),
\label{23}
\end{equation}
где интеграл $\widetilde I(\widetilde G)$ определяется формулой
{\rm (22)},  является базовой линейной комбинацией порядка $n$.}

{\bf Доказательство.} По условиям теоремы 2 всякий граф
$\widetilde{ G}$ из множества $\widetilde{\mathfrak G}_n^{(0)}$
принадлежит множеству $\widetilde{\mathfrak G}_n$. Отсюда 
по теореме 1 следует, что помеченное этим графом произведение
$\widetilde P_{1n}(\widetilde G)$ майеровских и больцмановских функций
является базовым произведением порядка $n$.
Следовательно, интеграл $\widetilde I(\widetilde G)$
является, по определению 9, базовым интегралом порядка $n$. Отсюда и
из условий теоремы 2 по определению 11 следует утверждение теоремы 2.
$\blacktriangleright$

{\bf Замечание 5.}  Для цели, поставленной в статье, нам достаточно
установить критерий сравнительной сложности представлений
ко\-эф\-фи\-ци\-ен\-та степенного
ряда лишь для случая, когда такие представления являются базовыми
линейными комбинациями, а сложность вычисления коэффициента при любом
из интегралов, входящих в такую линейную комбинацию, незначительна.
В дальнейшем такие базовые линейные комбинации мы будем называть
{\bf базовыми линейными комбинациями с коэффициентами незначительной
сложности}. $\blacksquare$

4. Предлагаемые критерии являются по существу критериями сложности
базовых линейных комбинаций с коэффициентами незначительной сложности.

Для начала сделаем следующее

{\bf Замечание 6.}  Из всех затрат машинного времени на вычисления, 
выполняемые для оценки базового интеграла, подавляющее большинство 
составляют затраты на вычисления значений майеровских и больцмановских 
функций, входящих в представление подынтегральной функции этого 
интеграла. Оставаясь в рамках самого грубого сравнения (так сказать, 
"в первом приближении"), можно считать, что из двух базовых интегралов
сложнее оценка того интеграла, в представление подынтегральной функции
  которого входит большее число майеровских и больцмановских функций. 
Если в представления подынтегральных функций обоих интегралов входит 
равное количество майеровских и больцмановских функций, то мы будем 
считать, что  оценки этих интегралов по сложности {\bf незначительно
отличаются} друг от друга, и будем говорить, что сложности этих оценок {\bf приблизительно
равны}. $\blacksquare$

Все предлагаемые в статье критерии опираются на содержащийся
в замечании 6 критерий сложности оценки базовых интегралов.

Простейшим таким критерием является длина $q(L)$ базовой линейной
комбинации $L$. Мы обозначим этот критерий $Cr_1$, полагая
$Cr_1(L) = q(L)$. Он применим тогда, когда сравниваемые базовые линейные
комбинации отличаются друг от друга по длине, тогда как входящие 
в них интегралы и их коэффициенты незначительно отличаются друг от друга
по своей сложности.

 В качестве другого критерия предлагается сумма
всех ребер всех графов-меток из множества ${\mathfrak G}(L)$, которые
маркируют интегралы, являющиеся членами данной
базовой линейной комбинации. Этот критерий обозначим через $Cr_2(L)$. Он
определяется формулой
\begin{equation}
Cr_2(L) = \sum_{G \in {\mathfrak G}(L)} (\left|X_f(G)\right| +
\left|X_{\widetilde f}(G)\right|),
\end{equation}
где  $\left|X_f(G)\right|$ --- мощность множества $X_f(G)$;
$\left|X_{\widetilde f}(G)\right|$ --- мощность множества
$X_{\widetilde f}(G)$ больцмановских функций.

Можно предложить еще  один, более точный, критерий. Он может быть применен
в~слу\-чае, когда для оценки каждого интеграла из базовой линейной
ком\-би\-на\-ции~$L$ порядка $n$, используется эквивалентная ему
вероятностная модель. В этой вероятностной модели оцениваемый интеграл
является математическим ожиданием произведения майеровских и больцмановских
функций от линейных комбинаций независимых случайных величин, принимающих
значения в $\nu$-мерном дейст\-ви\-тель\-ном евклидовом пространстве
${\bf R}^{\nu}$.
При этом каждая из этих случайных величин распределена с плотностью,
равной нормированному модулю майеровской функции. А число таких случайных
величин равно числу $n-1$. Таким образом, задача оценки базового интеграла,
маркируемого графом-меткой $G \in {\mathfrak G}_n$, сводится к оценке
математического ожидания произведения майеровских и больцмановских функций
от линейных комбинаций независимых непрерывных случайных величин.
В это произведение входят $\left|X_f(G)\right| - n + 1$ майеровских и
$\left|X_{\widetilde f}(G)\right|$ больцмановских функций.

Единственным известным способом оценки математического ожидания этого
произведения является построение аппроксимирующей дискретной
стохастической модели, которая получается из описанной выше
вероятностной модели заменой всех непрерывных
случайных величин аппроксимирующими их дискретными случайными величинами.
В результате задача оценки базового интеграла сводится к оценке
математического ожидания произведения майеровских и больцмановских функций
от линейных комбинаций независимых дискретных случайных величин.

При оценке этого математического ожидания в каждом статистическом испытании
приходится вычислять значения $\left|X_f(G)\right| - n + 1$ майеровских
и $\left|X_{\widetilde f}(G)\right|$ больцмановских функций. Из всех затрат машинного
времени на вычисления, выполняемые для оценки этого математического
ожидания,  подавляющее большинство составляют затраты на вычисления
значений майеровских и больцмановских
функций, число $N_1(G)$ которых определяется формулой
\begin{equation}
N_1(G) = \left|X_f(G)\right| - n + 1 + \left|X_{\widetilde f}(G)\right|.
\label{25}
\end{equation}
Поэтому это число может служить
{\bf модернизированным критерием сложности оценки базового интеграла},
маркируемого графом $G$.

О\,п\,р\,е\,д\,е\,л\,е\,н\,и\,е 16. В случае $N_1(G) = 0$ будем говорить,
что сложность оценки интеграла, маркируемого графом $G$,
по модернизированному критерию сложности оценки базового интеграла
является {\bf незначительной}. $\blacksquare$

{\bf Пример 2.} Рассмотрим граф $G = G(V_3; X_f, X_{\widetilde f})$, где
$X_f = \{\{1,2\}, \{2, 3\}\}$, \mbox{$X_{\widetilde f} = \varnothing$}.
Граф $G$ по определению принадлежит множеству ${\mathfrak G}_3$. Так как
его подграф $R(G) = G$  является связным, то помеченное графом $G$
каноническое произведение $P_{1n}(G)$, где $n = 3$, является
по определению 8 базовым.
Отсюда по определению 9 следует, что маркируемый графом $G$
интеграл $I(G)$, определенный формулой (16) является базовым интегралом
порядка 3.

Из определения множеств $X_f$ и $X_{\widetilde f}$ следует:
$\left|X_f\right| = 2$, $\left|X_{\widetilde f}\right| = 0$. Отсюда
по формуле (25) получаем
\begin{equation}
N_1(G) = 0.
\label{26}
\end{equation}
Из (26) по определению 16 следует вывод: сложность оценки интеграла,
маркируемого графом $G$, по модернизированному критерию сложности
$N_1(G)$ является незначительной. $\blacktriangleright$

Предлагаемый, более точный, критерий сложности базовой линейной
комбинации $L$ обозначим $Cr_3(L)$. Он опирается на критерий
сложности
$N_1(G)$ оценки базовых интегралов. В качестве такого критерия
предлагается
сумма по всем интегралам, входящим в данную базовую линейную комбинацию,
оценок сложности этих интегралов. Он определяется формулой

\begin{equation}
Cr_3(L) = \sum_{G \in {\mathfrak G}(L)}N_1(G),
\label{27}
\end{equation}
где $N_1(G)$ определяется формулой (25).

О\,п\,р\,е\,д\,е\,л\,е\,н\,и\,е 17. Пусть $L$ и $L_1$ --- две базовые
линейные комбинации с коэффициентами незначительной сложности. Будем
считать, что {\bf по критерию $Cr_i$,} где $i = 1, 2, 3$, {базовая 
линейная комбинация $L_1$ значительно сложнее базовой линейной 
комбинации $L$}, если $Cr_i(L_1) > Cr_i(L)$.
Если же \mbox{$Cr_i(L_1) = Cr_i(L)$}, то будем считать, что
по критерию $Cr_i$ сложность одной из этих двух базовых линейных 
комбинаций {\bf незначительно отличается} от сложности другой из них, и 
говорить что по критерию $Cr_i$ сложность одной из них 
{\bf приблизительно равна} сложности другой.

Если известно, что базовая линейная комбинация $L_1$ сложнее базовой
линейной комбинации $L$, и $Cr_i(L_1) = Cr_i(L)$, то будем считать, что
по критерию $Cr_i$ линейная комбинация $L_1$ {\bf незначительно 
сложнее} линейной комбинации $L$. $\blacksquare$

Предлагаемые критерии сложности базовых линейных комбинаций с 
коэффициентами незначительной сложности построены так,
чтобы они в основном, за некоторыми исключениями, удовлетворяли принципу:
если по данному критерию одна из двух базовых линейных комбинаций
зна\-чи\-тель\-но сложнее другой, то и на самом деле оценка значения
представляемой ею величины зна\-чи\-тель\-но
сложнее, чем оценка значения величины, представляемой другой базовой
линейной комбинацией. А в случае, когда по данному критерию сложность 
одной из двух базовых линейных комбинаций незначительно отличается 
от сложности другой из них, то и на самом деле сложность оценки значения величины,
представляемой одной из этих двух базовых линейных комбинаций, незначительно 
отличается от сложности оценки значения величины, представляемой другой базовой 
линейной комбинацией.

В~случае, когда выводы, сделанные на
основании значений одного из~критериев находятся в противоречии с выводами,
сделанными на осно\-ва\-нии
значений другого, более точного, критерия, предпочтение следует отдать
выводам, сделанным на~осно\-ва\-нии значений более точного критерия.

{\bf Пример 3.} Пусть $L$ и $L_1$ --- две линейные комбинации. При этом
в линейную комбинацию $L_1$ входят два интеграла, маркируемые графами $G$ и $G_1$.
Здесь $G$ --- граф, рассмотренный в примере 2, а граф
$G_1 = G_1(V_3; X_{f,1}, X_{{\widetilde f}, 1})$ имеет множество
майеровских ребер $X_{f,1} =  \{\{1,2\}, \{1, 3\}\}$ и множество
больцмановских ребер $X_{{\widetilde f},1} =  \{\{2, 3\}\}$. В линейную
комбинацию $L$ входят только один интеграл, маркируемый графом $G_1$.
При этом в обеих линейных комбинациях коэффициенты при базовых
интегралах $I(G)$ и $I(G_1)$ определены и равны 1.

Граф $G_1 = G_1(V_3; X_{f,1}, X_{{\widetilde f}, 1})$ по определению
принадлежит множеству ${\mathfrak G}_3$. Так как его подграф $R(G_1)$
является связным, то помеченное графом $G_1$ каноническое
произведение $P_{13}(G_1)$ является по определению 8 базовым. Отсюда
по определению~9 следует, что маркируемый
графом $G_1$ интеграл $I(G_1)$, определенный формулой~(16), где $n = 3$,
является базовым порядка 3.

В линейную комбинацию $L$ входит только один интеграл, маркируемый
графом $G_1$. Так как этот интеграл является базовым, а коэффициент при
нем задан и поэтому вообще не требуется никаких усилий для его вычисления,
то линейная комбинация $L$ является по определению 11 базовой линейной
комбинацией порядка $3$. А по замечанию 5 эта линейная комбинация
является базовой линейной комбинацией порядка $3$ с коэффициентами
незначительной сложности.

В примере 2 было доказано, что маркируемый графом $G$ интеграл
является базовым. Таким образом, оба входящих в линейную
комбинацию $L_1$ интеграла являются базовыми, а коэффициенты при них
заданы и поэтому вообще не требуется никаких усилий для их вычисления.
Отсюда следует, что линейная комбинация $L_1$ по определению 11 и
по замечанию 5 также является базовой линейной комбинацией порядка $3$
с коэффициентами незначительной сложности. Итак, базовая линейная
комбинация $L_1$ кроме интеграла, помеченного графом $G_1$, содержит еще
один базовый интеграл,
помеченный графом~$G$. Естественно полагать, что базовая
линейная комбинация~$L_1$ сложнее, чем базовая линейная комбинация~$L$.

Используя определение критерия сложности оценки базового интеграла
формулой (25), найдем значение этого критерия для интеграла, помеченного
графом~$G_1$:
\begin{equation}
N_1(G_1) = \left|X_f(G_1)\right| - n + 1 +
\left|X_{\widetilde f}(G_1)\right| = 1.
\label{28}
\end{equation}
Значение этого критерия для интеграла, помеченного
графом $G$, было найдено в~примере~2 (см. формулу (26)).

Исходя из определения критерия $Cr_3$ формулой (27) и используя
формулы (26) и (28), найдем значения этого
критерия для базовых линейных комбинаций $L$ и $L_1$:
\begin{equation}
Cr_3(L) = Cr_3(L_1) = 1.
\label{29}
\end{equation}

Из формулы (29) по определению 17 следует, что
базовая линейная комбинация $L_1$ по критерию $Cr_3$ незначительно сложнее базовой
линейной комбинации $L$. $\blacktriangleright$

Из определения критерия $Cr_3$ и определения 17 вытекает

{\bf Следствие 3.} {\it Пусть $L$ и $L_1$ --- две базовые линейные комбинации
с коэффициентами незначительной сложности, удовлетворяющие условиям:

{\rm 1.} Длина линейной комбинации $L_1$ больше длины линейной
комбинации $L$.

{\rm 2.} Каждый интеграл, входящий в линейную комбинацию $L$, входит
и в линейную комбинацию $L_1$.

Допустим, что среди интегралов, входящих в линейную комбинацию $L_1$ и
не вхо\-дя\-щих в линейную комбинацию $L$, имеется хоть один интеграл,
имеющий ненулевое значение критерия сложности $Cr_3$ его оценки. Тогда
по критерию $Cr_3$ базовая ли\-ней\-ная комбинация $L_1$ значительно
сложнее базовой линейной комбинации $L$. В противном случае базовая
линейная комбинация $L_1$ незначительно сложнее базовой линейной
ком\-би\-на\-ции $L$.}

О\,п\,р\,е\,д\,е\,л\,е\,н\,и\,е 18. Базовое произведение $P(G)$
называется {\bf полным}, если его граф-метка $G$ является полным.
В противном случае базовое произведение называется 
{\bf не\-пол\-ным}. $\blacksquare$

О\,п\,р\,е\,д\,е\,л\,е\,н\,и\,е 19. Базовый интеграл называется
{\bf полным}, если его подынтегральная функция является полным базовым
произведением. Базовый интеграл называется {\bf неполным}, если
его подынтегральная функция является неполным базовым 
произведением. $\blacksquare$

О\,п\,р\,е\,д\,е\,л\,е\,н\,и\,е 20. Базовая линейная комбинация
называется {\bf полной}, если все входящие в нее интегралы являются
полными. В противном случае базовая линейная комбинация называется
{\bf неполной}. $\blacksquare$

Из определения представлений Ри-Гувера [47] и определений 18,
19 и 20 вытекает

{\bf Следствие 4.} {\it При любом $n > 1$ представление Ри-Гувера
вириального коэффициента $B_n$ является полной базовой линейной
комбинацией порядка $n$ с коэффициентами незначительной сложности}.

Из определений 18 и 19 и замечания 6 вытекает следующее

{\bf Замечание 7.}  Пусть один из двух базовых интегралов является
полным, а другой --- неполным, причем оба этих интеграла маркируются
графами с одним и тем же множеством вершин. Тогда оценка полного
интеграла значительно сложнее, чем оценка неполного интеграла. 
$\blacksquare$

Из замечания 7 вытекает

{\bf Следствие 5} {\it Пусть $L_1$ --- неполная базовая линейная комбинация
порядка $n$ с коэффициентами незначительной сложности, а $L_2$ --- полная
базовая линейная комбинация порядка $n$ с коэффициентами незначительной
сложности. Если число интегралов в~линейной комбинации $L_1$ не больше
числа интегралов в~линейной комбинации $L_2$, то, по замечанию 7 и
критериям $Cr_2$ и $Cr_3$, линейная комбинация $L_2$ значительно сложнее
линейной комбинации $L_1$.}

5. В рамках метода каркасных сумм можно выделить два подхода.

Для изложения первого из них нам потребуется ввести определение древесной
суммы. С целью упрощения изложения и не стремясь к максимальной общности, 
мы дадим это определение в смысле, хотя и не самом общем, но достаточном 
для целей, поставленных в этой статье.

Для этого введем следующие определения:

$T_n = \{t\}$ --- множество всех помеченных деревьев с множеством вершин
$V_n$, где $n > 1$, и корнем $1$;

$X_f(t) = \{\{u,v\}\}$ --- множество ребер дерева $t \in T_n$;

$\XX_{ad}(t)$ --- множество допустимых ребер [10, 13, 17]
  дерева $t \in T_n$;

\begin{equation}
I(t) =
\int_{(\R)^{n-1}}\prod\limits_{\{u,v\}\in X_f(t)}f_{uv}
\prod\limits_{\{\widetilde u,\widetilde v\}\in \XX_{ad}(t)}
(1+f_{\widetilde u\widetilde v})(d{\bf r})_{1,n-1},
\label{30}
\end{equation}
где $t \in T_n$.

Пусть $T'$ --- непустое подмножество множества деревьев $T_n$,
где $n > 1$; а каждому дереву $t \in T'$ поставлено в соответствие
множество допустимых ребер $\XX_{ad}(t) = \{\{u,v\}\}$.
Обозначим $L(T')$ линейную комбинацию интегралов, определенную формулой
\begin{equation}
L(T') = \sum_{t\in T'}c(t)I(t),
\label{31}
\end{equation}
в которой при каждом $t \in T'$ интеграл $I(t)$ определен формулой
(30), а коэффициент $c(t)$ при интеграле $I(t)$ является
определенной на множестве деревьев $T'$ действительной функцией.

О\,п\,р\,е\,д\,е\,л\,е\,н\,и\,е 21. Линейная комбинация интегралов 
$L(T')$, определенная формулой (31), в которой $T' \subset T_n$ 
и $n \ge 2$, называется {\bf древесной суммой}. $\blacksquare$

{\bf Замечание 8.}  Из определения множества допустимых ребер
$\XX_{ad}(t)$  вытекает, что это множество
не пересекается с множеством $X_f(t)$ ребер дерева
$t \in T_n$ и состоит из попарно различных ребер, каждое из которых
соединяет две несмежные вершины дерева $t$. $\blacksquare$

{\bf Теорема 3.} {\it Древесная сумма, определенная формулой
{\rm (31)},
где $T' \subset T_n$ и \mbox{$n > 1$}, является базовой линейной
комбинацией порядка $n$, а каждое дерево $t \in T'$ --- укомплектованным
графом-меткой интеграла $I(t)$, определенного формулой {\rm (30)}.
При этом каждому такому дереву $t$ поставлено в соответствие, в качестве
дополняющего множества, множество
допустимых ребер $\XX_{ad}(t) = \{\{u,v\}\}$.}

{\bf Доказательство.} Определение интеграла $I(t)$ формулой (30)
означает, что для каждого дерева $t \in T'$ определено поставленное ему
в соответствие конечное множество $\XX_{\rm ad}(t)$ допустимых ребер.
По замечанию 8 это множество не пересекается с множеством $X_f(t)$ и
состоит из попарно различных ребер, каждое
из которых соединяет две несмежные вершины дерева $t$. 
Из определения интеграла $I(t)$ формулой (30) следует,
что подын\-тегральная функция этого интеграла является произведением
майеровских и больцмановских функций. При этом множество ребер
$X_f(t)$ дерева $t$, маркирующего интеграл $I(t)$, определяет
множество $F$ всех майеровских функций этого произведения и является, по
определению 3, множеством майеровских ребер по отношению к множеству $F$
майеровских функций. А по определению 4 множество допустимых ребер
$\XX_{ad}(t)$ определяет множество $\widetilde F$ всех больцмановских
функций этого произведения и является множеством больцмановских
ребер по отношению к множеству $\widetilde F$ больцмановских функций.
Таким образом, множество $\XX_{ad}(t)$ является дополняющим множеством
дерева $t$ по определению дополняющего множества, а 
множества $X_f(t)$ и $\XX_{ad}(t)$ образуют упорядоченную
пару ${\bf X} = (X_f, X_{\widetilde f})$.

По определению множества деревьев $T_n$ всякое
дерево $t\in T_n$ является связным графом с множеством вершин $V_n$ и,
значит, имеет место равенство $V(X_f(t)) = V_n$. Отсюда и из замечания 8
следует равенство $V(X_f)\bigcup V(X_{\widetilde f}) = V_n$. Из этого
равенства по определению 5 следует, что упорядоченная пара множеств
${\bf X} = (X_f, X_{\widetilde f})$ является канонической. 

Из полученных результатов следует, что  всякое дерево $t\in T_n$ 
принадлежит множеству $\widetilde{\mathfrak G}_n$ по его определению.
Отсюда следуют выводы: 
1) по тереме 1 каждое дерево $t \in T'$ является 
укомплектованным графом-меткой интеграла $I(t)$, определенного 
формулой {\rm (30)}; 2) по теореме 2 определенная формулой
{\rm (31)} древесная сумма является является базовой линейной
комбинацией порядка $n$. Теорема 3 полностью доказана. 
$\blacktriangleright$

6. В качестве примера представления коэффициентов степенных рядов древесными
суммами можно привести полученные автором [3, 10, 17] представления
майеровских коэффициентов $b_n$, свободные от асимптотической катастрофы.
Вначале были получены [3] такие представления, в которых коэффициент
$b_n$ выражался произведением числа $1/n!$ на сумму всех интегралов,
маркируемых $n$-вершинными корневыми помеченными деревьями
[25, 28, 10, 17] с корневой вершиной, помеченной числом $1$.
При этом каждому маркирующему дереву $t$ было поставлено в соответствие
множество допустимых ре\-бер ${\XX_{ad}(t) = \{\{u,v\}\}}$.
По определению 21 такая сумма является дре\-вес\-ной суммой. В этой сумме
коэффициент при каждом входящем
в нее интеграле равен единице. Поэтому не требуется никаких вычислений
для определения значений коэффициентов при входящих в эту сумму
интегралах.
Отсюда по теореме 3 и по замечанию 5 следует, что эта древесная
сумма является базовой линейной комбинацией порядка $n$ с коэффициентами
незначительной сложности.

Впоследствии эти представления были упрощены [10, 17]. 
С этой целью было
вве\-де\-но бинарное отношение максимального изоморфизма корневых
помеченных деревьев. Это отношение обладает свойствами рефлексивности,
симметричности и тран\-зи\-тив\-но\-с\-ти, то есть является отношением
эквивалентности [21] и разбивает множество $\{T_n\}$, состоящее из всех
помеченных деревьев с множеством вершин $V_n = \{1, 2,\ldots,n\}$
и корнем $1$ на классы максимально изоморфных деревьев.
Эти классы обладают очень полезным свойством: в вышеупомянутом
представлении майеровских коэффициентов равны все интегралы, входящие
в сумму, представляющую коэффициент $b_n$, и маркируемые максимально
изоморфными деревьями. Было найдено кон\-с\-т\-рук\-тив\-ное определение
 такого подмножества $TR(n)\subset T_n$ [10, 17], в котором
никакие два дерева не являются максимально изоморфными, а мощность
которого равна числу классов максимально изоморфных деревьев,
принадлежащих множеству $T_n$.

Используя  представления коэффициента $b_n$ суммой всех интегралов,
мар\-ки\-ру\-е\-мых $n$-вершинными помеченными деревьями с корневой 
вершиной, помеченной числом $1$, разложение множества корневых
помеченных деревьев на классы максимально изоморфных деревьев и 
вышеупомянутое свойство максимально изоморфных деревьев, удалось получить
представления майеровских коэффициентов $b_n$ в виде:
\begin{equation}
b_n =
\frac{1}{n!}\sum_{t\in T\!R(n)}\left|TI(t)\right|I(t).
\label{32}
\end{equation}
Здесь $TI(t)$ --- совокупность деревьев, принадлежащих множеству $T_n$ и
максимально изоморфных дереву $t$;

$\left|TI(t)\right|$ --- мощность множества $TI(t)$;

$I(t)$ --- интеграл, определенный формулой (30).

Число деревьев во множестве $TI(t)$ полностью определяется деревом $t$
по формуле
\begin{equation}
\left|TI\,(t)\right|=
(n - 1)!\,\Bigl(
\prod_{i=1}^{H(t)-1}n(t,i)!\Bigr)^{-1}
\Bigl(\prod_{i=1}^{n(t,H(t)-1)}(d(t,i) - 1)!
\Bigr)^{-1}.
\label{33}
\end{equation}
Здесь $H(t)$ --- высота [10, 17, 5] дерева $t$;
$n(t,i)$ --- число вершин дерева $t$, находящихся на высоте $i$;
$d(t,i)$ ---степень $i$-ой вершины из множества всех вершин дерева $t$, находящихся
на высоте $H(t) - 1$.

{\bf Лемма 3.} {\it Представление майеровского коэффициента $b_n$
формулами {\rm (32)} и {\rm (30)} при $n > 1$ является
базовой линейной комбинацией порядка $n$ с коэффициентами незначительной
 сложности.}

{\bf Доказательство.} Сумма в правой части равенства (32) имеет
следующие свойства: 1) множество деревьев $TR\,(n)$ является подмножеством
множества $T_n$; 2) интегралы,  входящие в эту сумму, определяются формулой
(30); 3) коэффициентом при каждом из этих мнтегралов является число
деревьев, максимально изоморфных дереву $t$, маркирующему этот интеграл;
это число определяется деревом $t$ по формуле (33).
Отсюда по определению 21 следует, что эта сумма является древесной суммой.
По теореме 3 эта древесная сумма является базовой линейной комбинацией
по\-ряд\-ка~$n$.

Из определения коэффициентов этой древесной суммы формулой (33)
следует, что сложность вычисления этих коэффициентов незначительна.
Поэтому эта древесная сумма является базовой линейной комбинацией
порядка $n$ с коэффициентами  незначительной  сложности. Лемма 
доказана. $\blacktriangleright$

Число деревьев во множестве $TR(n)$ вычисляется по формуле
\begin{multline}
\left|TR\,(n)\right|=1 + \left(2^{n - 2} - 1\right)+{}\\
+\sum_{H=3}^{n - 1} \,
\sum_{{\bf n} \in {\bf N}(H,\,\, n - 1)}\;
\frac{(n(H - 1) + n(H) - 1)!}{n(H)!(n(H-1)-1)!}
\prod_{i=2}^{H-1}\{[n(i-1)]^{n(i)}\}.
\label{34}
\end{multline}
Здесь ${\bf N}(H, k) = \{(n(1),n(2), \ldots, n(H))\}$ --- множество $H$-мерных
векторов, компонентами которых являются натуральные числа, а сами векторы
удовлетворяют условию: $\sum\limits_{i=1}^H n(i) = k$.

Вычисления по формуле (34) приведены в таблице 1. В этой таблице
приведены длины базовых линейных комбинаций, являющихся представлениями
майеровских коэффициентов $b_n$ в виде древесных сумм
по формулам (32) и (30).

Сравним теперь сложность этих представлений со сложностью представлений
Ри-Гувера вириальных коэффициентов.

В простейшем случае, когда $n = 2$, и майеровский коэффициент $b_2$, и
вириальный коэффициент $B_2$ представляются через один и тот же интеграл
и их представления отличаются лишь знаком. Упрощать здесь нечего.

Далее, из таблицы 1 явствует, что при $n = 7, 8, 9, 10$ представление
майеровского коэффициента~$b_n$ по формуле (32) содержит меньшее 
число слагаемых интегралов, чем представление Ри-Гувера вириального
коэффициента $B_n$. Следовательно, при этих значениях $n$ по критерию
$Cr_1$
сложность представления Ри-Гу\-вера вириального коэффициента $B_n$
значительно больше, чем сложность представления майеровского коэффициента
$b_n$ древесной суммой по формулам (32) и (30).

Посмотрим теперь, какой результат получается по критериям $Cr_2$ и $Cr_3$.

Из определения множества
$\XX_{ad}(t) = \{\{u,v\}\}$ допустимых ребер дерева $t$ следует, что
при любом $n > 2$ определенная формулами (32) и (30) древесная
сумма удовлетворяет условию: в этой сумме только  один интеграл,
маркируемый звездой [25, 28],все ребра которой инцидентны ее корню, 
является полным базовым интегралом; а все остальные
интегралы в этой сумме являются неполными базовыми интегралами.
Отсюда по определению 20 и лемме 3 следует, что при любом $n > 2$
представление майеровского коэффициента $b_n$ древесной суммой
по формулам (32) и (30) является неполной базовой линейной
комбинацией порядка $n$ с коэффициентами незначительной  сложности.

С другой стороны, по следствию 4 представление Ри-Гувера вириального
коэффициента $B_n$ является полной базовой линейной комбинацией
порядка $n$ с коэффициентами незначительной сложности.

 Из вышеизложенного по следствию 5 вытекает, что при значениях
$n = 7, 8, 9, 10$ по критериям $Cr_2$ и $Cr_3$ сложность представления
Ри-Гу\-вера вириального коэффициента $B_n$ значительно больше, чем
сложность представления майеровского коэффициента $b_n$ древесной суммой
по формулам (32) и (30).

Заметим, что при $n = 8, 9, 10$ число интегралов в сумме, представляющей
по методу Ри-Гувера вириальный коэффициент $B_n$ сильно превышает число
интегралов в сумме, представляющей майеровский коэффициент $b_n$
по формулам (32) и (30). Поэтому по следствию 5 при этих
значениях $n$ представление майеровского коэффициента $b_n$
по формулам (32) и (30) является значительно более
простым, чем представление вириального коэффициента $B_n$ по методу
Ри-Гувера.

Однако, при $n = 3, 4, 5, 6$ сравниваемые представления не удовлетворяют
условиям следствия 5. Значит, при этих значениях $n$ данное следствие
невозможно применить для такого сравнения. А по критерию $Cr_1$
представление
майеровского коэффициента $b_n$ по формулам (32) и (30)
является более сложным, чем представление вириального коэффициента $B_n$
по методу Ри-Гувера. Применим более точный критерий $Cr_3$ для проверки
этого вывода, сделанного по самому простому критерию $Cr_1$.

В таблице 3 приведены вычисленные при $n = 3, 4, 5, 6$ значения этого
критерия для представлений вириальных коэффициентов $B_n$ по методу
Ри-Гувера и базовых линейных комбинаций, являющихся представлениями
майеровских коэффициентов $b_n$ в виде древесных сумм
по формулам (32) и (30). Используя эти значения
и применяя определение 17, получаем следующие результаты:

1. сложность представления майеровского коэффициента $b_3$ древесной суммой
по формулам (32) и (30) приблизительно равна сложности представления
Ри-Гувера вириального коэффициента $B_3$;

2. представления майеровских коэффициентов $b_4$ и $b_5$ древесными
суммами по формулам (32) и (30) значительно сложнее
представлений Ри-Гувера вириальных коэффициентов $B_4$ и $B_5$
соответственно;

3. представление Ри-Гувера вириального коэффициента $B_6$ значительно
сложнее, чем представления майеровского коэффициента $b_6$ древесной
суммой по формулам (32) и (30).

7. Другим примером представления коэффициентов степенных рядов в виде
древесных сумм является представление коэффициентов $a_n$ разложения
отношения активности $z$ [22, 24, 43, 48] к плотности $\varrho(z)$ в ряд
по степеням активности $z$:
\begin{equation}
z/\varrho(z) =1-
\sum\limits_{n=2}^{\infty}n a_n z^{n-1}.
\label{35}
\end{equation}
Это разложение рассматривалось Либом [40] и Пенроузом [44].

Пенроузом было предложено два метода нахождения коэффициентов $a_n$: либо
весьма сложным путем  с помощью уравнений Кирквуда--Зальцбурга;
либо более простым путем,  исходя из соотношений
\begin{equation}
nb_n=\sum_{q=1}^{n -1}(q+1)a_{q+1}(n-q)b_{n-q}
\label{36}
\end{equation}
 между этими
коэффициентами и майеровскими коэффициентами $b_n$.

В работах [4, 31, 10, 17] было предложено представление коэффициентов
$a_n$ в виде суммы интегралов, маркируемых деревьями. С этой целью было
определено множество $T(n,0)$, состоящее из всех деревьев множества $T_n$,
удовлетворяющих условиям:

а) любой слой этого дерева, за исключением нулевого и, может быть, последнего,
состоит не менее чем из двух вершин;

 б) за исключением нулевого слоя, дерево
не имеет ни одного слоя, в котором лишь самая старшая вершина имеет степень,
большую чем единица.

Это позволило получить [4, 31, 10, 17] свободные от асимптотической катастрофы
представления коэффициентов $a_n$
в виде суммы всех интегралов, маркируемых деревьями из множества $T(n,0)$:

\begin{equation}
a_n = \frac{1}{n!}\sum_{t\in T_n}\,I(t),
\label{37}
\end{equation}
где $I(t)$ --- интеграл, определенный формулой (30).

Впоследствии эти представления были упрощены [10, 17]. 
С этой целью было определено множество $TR(n,0) = TR(n) \cap T(n,0)$ 
[10, 17].

Из определения отношения максимального изоморфизма корневых помеченных
деревьев и определения множеств $T(n,0)$ и $TR(n,0)$ следует, что множество
$T(n,0)$ разлагается на~клас\-сы $TI(t)$ максимально изоморфных деревьев,
маркируемых деревьями из множества $TR(n,0)$. А множество $TR(n,0)$ состоит
из всех деревьев $t$, являющихся метками входящих во множество $T(n,0)$
классов $TI(t)$ максимально изоморфных деревьев.

Используя представление коэффициентов $a_n$ формулой (37),  понятие
максимального изоморфизма корневых помеченных деревьев, разложение
множества $T(n,0)$ на классы максимально изоморфных деревьев и свойства
максимально изоморфных деревьев, автором были предложены более простые
представления коэффициентов $a_n$, свободные от асимптотической катастрофы:
\begin{equation}
a_n = \frac{1}{n!}\sum_{t\in T\!R(n,0)}\,\left|TI(t)\right|I(t).
\label{38}
\end{equation}
Здесь, как и в формуле (32), $I(t)$ --- интеграл, определенный
формулой (30); $\left|TI(t)\right|$ --- число деревьев во множестве
$TI(t)$, определяемое деревом $t$ по формуле (33).

Число деревьев во множестве $TR(n,0)$ вычисляется по формуле
\begin{multline}
\left|TR\,(n,0)\right|=1+\MO{\sum\nolimits^{\LT{\prime}
}_{\LT{\scriptstyle 2}}}_{\bf n}\quad
\left(\frac{[n(2)+n(1)-1]!}{[n(1)-1]!\,n(2)!}- 1\right)+{}\\
{}\\
+\sum_{H=3}^N
\MO{\sum\nolimits^{\LT{\prime}}
_{\LT{\scriptstyle H}}}_{\bf n}\quad
\left(\frac{[n(H)+n(H-1)-1]!}{[n(H-1) - 1]!\,n(H)!} -1\right)
\prod_{i=2}^{H-1}\left([n(i-1)]^{n(i)} -1\right),
\label{39}
\end{multline}
где $N=\lceil n/2 \rceil$ --- наименьшее из тех целых чисел, которые
не меньше числа $n/2$, а сим\-вол
$\sum\limits_{\bf n}{\vphantom{\left(sum\right)'}}'_{\scriptscriptstyle H}$
\rule[-3ex]{0em}{3ex}
в формуле (39) означает суммирование по~всем $H$-мерным векторам
$(n_1,n_2,\ldots,n_{{}_H})$, компоненты которых являются натуральными
числами, а сами векторы удовлетворяют условиям:

\ \ \ \ а) $n_i \ge 2$, \quad $i=1,2,\ldots,H-1$; \qquad б) $n_{{}_H} \ge 1$; \qquad
в)~$\sum\limits_{i=1}^H n(i) = n-1$.

{\bf Лемма 4.} {\it Представление коэффициента $a_n$ формулами
{\rm (38)} и {\rm (30)} при $n > 3$ является базовой линейной
комбинацией порядка $n$ с коэффициентами незначительной  сложности.}

{\bf Доказательство.} Сумма в правой части равенства (38) имеет
следующие свойства: 1) множество деревьев $TR\,(n,0)$ является
подмножеством множества $T_n$;
2) интегралы, входящие в эту сумму, определяются формулой (30);
3) коэффициентом при каждом из этих мнтегралов является число деревьев,
максимально изоморфных дереву $t$, маркирующему этот интеграл; это число
определяется деревом $t$ по формуле (33). Отсюда по определению 21
следует, что эта сумма является древесной суммой.

По теореме 3 эта древесная сумма является базовой линейной комбинацией
по\-ряд\-ка~$n$.

Из определения коэффициентов этой древесной суммы формулой (33)
следует, что сложность вычисления этих коэффициентов незначительна. Поэтому эта
древесная сумма является базовой линейной комбинацией порядка $n$
с коэффициентами незначительной  сложности. Лемма доказана.
$\blacktriangleright$

{\bf Замечание 9.}  Из определения [4, 31, 10, 17] множества
$\XX_{ad}(t) = \{\{u,v\}\}$ до\-пус\-ти\-мых ребер дерева $t$ следует, что
при любом $n > 3$ определенная формулами (38) и (30) древесная
сумма удовлетворяет условию: в этой сумме только  один интеграл,
маркируемый звездой, все ребра которой инцидентны ее корню, является 
полным базовым интегралом; а все остальные
интегралы в этой сумме являются неполными базовыми интегралами.
Отсюда по определению 20 и лемме 4 следует, что при любом $n > 3$
представление коэффициента $a_n$ древесной суммой по формулам (38)
и (30) является неполной базовой линейной комбинацией порядка $n$
с коэффициентами незначительной сложности. $\blacksquare$

Из представления (32) майеровских коэффициентов $b_n$ и
из представления (38) коэффициентов $a_n$ очевидно, что
$b_2 = a_2$. Указанные представления этих коэффициентов совпадают и
имеют одну и ту же сложность.

А из определения множеств $TR(n)$ и $TR(n,0)$ при $n >2$ следует, что 
множество $TR(n,0)$ является собственным подмножеством множества $TR(n)$.
Отсюда вытекают два следствия:

1. При любом $n > 2$ длина базовой линейной комбинации, представляющей
майеровский коэффициент $b_n$ по формулам (32) и (30), больше
длины базовой линейной комбинации, представляющей коэффициент $a_n$
по формулам (38) и (30).

2. При любом $n > 1$ каждый интеграл, входящий в сумму, представляющую
по формулам (38) и (30) коэффициент $a_n$, входит и в сумму,
представляющую по формулам (32) и (30) майеровский коэффициент
$b_n$.

Из определения множества деревьев $TR(n)$ следует, что
множество $TR(3)$ состоит из двух деревьев, являющимися графами $G$ и $G_1$,
введенными в примерах 2 и 3 соответственно. Далее, из определения множества
деревьев $TR(n,0)$ следует, что множество
$TR(3, 0)$ состоит
из одного дерева, являющегося графом $G_1$. Из~ре\-зуль\-та\-тов, полученных
в примере 3, явствует, что базовая линейная комбинация, представляющая
майеровский коэффициент $b_3$ по формулам (32) и (30),
незначительно сложнее, чем базовая линейная комбинация, представляющая
коэффициент $a_3$ по формулам (38) и (30).

При $n > 3$ множество $T_n$ содержит по крайней мере одно дерево, которое
не принадлежит множеству $T(n, 0)$ и имеет непустое множество допустимых
ребер. К таким деревьям относятся, в частности, все деревья
из множества $T_n$ высоты $H > 1$, не являющиеся цепью и имеющие такой
слой вершин высоты меньшей, чем $H$, в котором лишь самая старшая вершина
имеет степень большую чем два.
Очевидно, что интегралы, маркируемые такими деревьями, имеют положительное
значение критерия $Cr_3$ сложности их оценки. Они входят в базовую линейную
комбинацию, представляющую майеровский коэффициент $b_n$
по формулам (32) и (30), и не входят в базовую линейную
комбинацию, представляющую коэффициент $a_n$ по формулам (38)
и (30).

Таким образом, в рассматриваемой ситуации удовлетворяются все условия
след\-с\-т\-вия~2. Отсюда по следствию 3 следует вывод, что при $n > 3$
представление базовой линейной комбинацией майеровского коэффициента 
$b_n$ по формулам (32) и (30) значительно сложнее, чем представление
базовой линейной комбинацией коэффициента $a_n$ по формулам (38) 
и (30).

В таблице 3 приведены вычисленные при $n = 3, 4, 5, 6$ значения критерия
$Cr_3$ для базовых линейных комбинаций, являющихся представлениями
майеровских коэффициентов $b_n$ в виде древесных сумм по формулам
(32) и (30), и для базовых линейных комбинаций, являющихся
представлениями коэффициентов $a_n$ в виде древесных сумм по формулам
(38) и (30).

Эти значения являются численным подтверждением полученной теоретическим путем
сравнительной оценки сложности этих базовых линейных комбинаций.

Отсюда следует вывод, что прямой метод оценки коэффициентов
$a_n$, основанный на их представлении древесными суммами по формулам
(38) и (30), проще и рациональнее предложенного Пенроузом
метода оценки коэффициентов $a_n$, исходящего из соотношений (36),
между этими коэффициентами и майеровскими коэффициентами $b_n$.
Соотношения же (36) целесообразнее использовать для представления
коэффициентов $b_n$ через коэффициенты $a_n$, чтобы затем применить
эти представления как для оценки майеровских коэффициентов $b_n$, так и
для оценки вириальных коэффициентов $B_n$.

Вычисления по формуле (39) при $n = \overline{2, 10}$ приведены
в таблице 1. Данные этой таблицы подкрепляют уже сделанные выводы. 
По этим данным при $n = \overline{4, 10}$  число интегралов в сумме, 
представляющей
майеровский коэффициент $b_n$ по формуле (32) более чем в $2$ раза
больше числа интегралов
в представлении коэффициента $a_n$ по формуле (38).
Отсюда по простейшему критерию, каким является длина базовой линейной
комбинации, следует вывод: при $n = \overline{2, 10}$ такое представление
коэффициента $a_n$ в несколько раз проще, чем представление майеровского
коэффициента $b_n$ древесной суммой по~фор\-мулам (32) и (30).

Данные таблицы 1 дают также возможность сравнить сложность представления
коэффициента $a_n$ древесной суммой по формулам (38) и (30)
со сложностью представления вириального коэффициента $B_n$ по методу
Ри-Гувера.

Рассмотрим простейший случай, когда $n = 2$. В этом случае представление
коэффициента $a_2$ по формуле (38) отличается от представления
вириального коэффициента $B_2$ по методу Ри-Гувера лишь знаком.
Очевидно, что сложность такого представления коэффициента $a_2$ равна
сложности представления вириального коэффициента $B_2$ по методу
Ри-Гувера.  Упрощать здесь нечего.

Рассмотрим случай $n = 3$. Представление коэффициента $a_n$ по формулам
(38) и (30) является полной базовой линейной комбинацией
порядка $3$ и содержит только один полный базовый интеграл. Также и
представление вириального коэффициента $B_3$ по методу Ри-Гувера является
полной базовой линейной комбинацией порядка $3$ и содержит только один
полный базовый интеграл. При значении $n = 3$ и критерий $Cr_1$,
и критерий $Cr_2$, и критерий $Cr_3$ принимают одинаковые значения как
для базовой линейной комбинации, представляющей коэффициент $a_3$,
так и для базовой линейной комбинации, представляющей вириальный
коэффициент $B_3$ по методу Ри-Гувера. Значит, по критериям $Cr_2$ и 
$Cr_3$ сложности этих представлений являются,
по определению 17, приблизительно равными и незначительно отличаются
друг от друга.

Из результатов вычислений, приведенных в таблице 1, явствует, что при
$n = 4, 5$ длина базовой линейной комбинации, представляющей
коэффициент $a_n$, равна длине базовой линейной комбинации, являющейся
представлением Ри-Гувера вириального коэффициента $B_n$. Значит, при этих
значениях $n$ сложность представления коэффициентов $a_n$ древесными
суммами по формулам (38) и (30) по критерию $Cr_1$
не отличается от сложности представления Ри-Гувера вириального
коэффициента $B_n$.

Из той же таблицы явствует, что при $n = \overline{6,10}$ длина базовой
линейной комбинации, представляющей коэффициент $a_n$, меньше длины базовой
линейной комбинации, являющейся представлением Ри-Гувера вириального
коэффициента $B_n$. Значит, при этих значениях $n$ по критерию $Cr_1$
сложность представления коэффициентов $a_n$ древесными суммами
по формулам (38) и (30)  меньше сложности представления
Ри-Гувера вириального коэффициента $B_n$.

Посмотрим, что дает применение критериев $Cr_2$ и $Cr_3$.
Из формулы (39) следует, что при $n > 3$ мощность множества
$TR(n, 0)$ больше единицы. Следовательно, при $n > 3$ длина древесной
суммы, представляющей коэффициент $a_n$, более единицы. Отсюда и
из определения множества $\XX_{ad}(t) = \{\{u,v\}\}$ допустимых
ребер дерева $t$ следует, что при $n > 3$ в древесной сумме,
представляющей коэффициент $a_n$, только один интеграл, маркируемый
звездой, все ребра которой инцидентны ее корню, является полным 
базовым интегралом; а все остальные интегралы
в этой сумме являются неполными базовыми интегралами. Отсюда
по замечанию 9 следует, что при $n > 3$ древесная сумма, представляющая
коэффициент $a_n$, является неполной базовой линейной комбинацией
порядка $n$ с коэффициентами незначительной сложности.
С другой стороны, по следствию 4 представление Ри-Гувера вириального
коэффициента $B_n$  является полной базовой линейной комбинацией
порядка $n$ с коэффициентами незначительной сложности.

Из результатов вычислений, приведенных в таблице 1, явствует, что при
$n = \overline{4,10}$ длина базовой линейной комбинации, представляющей
коэффициент $a_n$, не больше длины базовой линейной комбинации,
являющейся представлением Ри-Гувера вириального коэффициента $B_n$.
Значит, при этих значениях $n$, по следствию 5, представление
коэффициентов $a_n$ древесными суммами по формулам (38) и
(30) по критериям $Cr_2$ и $Cr_3$ значительно проще, чем
представлением Ри-Гувера вириального коэффициента $B_n$.

При $n = 7, 8$ число интегралов в сумме, представляющей коэффициент $a_n$
древесной суммой по формуле (38),  значительно меньше
 числа интегралов в сумме, представляющей вириальный коэффициент $B_n$
по методу Ри--Гувера. Отсюда следует вывод:  представление
вириального коэффициента $B_n$ по методу Ри-Гувера  при
$n = 7, 8$ в разы сложнее, чем представление коэффициента $a_n$ древесной
суммой по формуле (38) при том же значении $n$.

При \mbox{$n = 9, 10$} число интегралов в сумме, представляющей
вириальный коэффициент $B_n$ по методу Ри--Гувера, весьма  значительно
превышает число интегралов в сумме, представляющей коэффициент $a_n$
дре\-вес\-ной суммой по формуле (38). Отсюда следует вывод:
представление вириального коэффициента $B_n$ по методу Ри-Гувера при
\mbox{$n = 9, 10$} во много раз сложнее, чем представление коэффициента
$a_n$ древесной суммой по формуле (38) при том же значении $n$.

8. Еще одним примером успешного применения метода каркасных сумм являются
полученные этим методом представления вириальных коэффициентов.
В рамках этого метода были разработаны два способа представления
вириальных коэффициентов.

Первый из них состоит в следующем: каждый вириальный коэффициент
представляется в виде многочлена от древесных сумм. В качестве примеров
такого способа представления вириальных коэффициентов можно привести
представления вириальных коэффициентов, свободные от асимптотической
катастрофы, двух видов:
1) в виде многочленов от древесных сумм, представляющих майеровские
коэффициенты $b_n$ и \linebreak
2) в виде многочленов от древесных сумм, представляющих коэффициенты
$a_n$.

Представления вириальных коэффициентов в виде многочленов от древесных
сумм, представляющих майеровские коэффициенты $b_n$ можно получить,
используя ре\-зуль\-та\-ты, полученные Майером [22, 41, 42, 43].
В статье [43]  дано представление (в виде многочленов от майеровских
коэффициентов $b_n$)
величин $\beta_{\mu}$, через которые выражаются вириальные коэффициенты
по формуле
\begin{equation}
B_n = {\frac{n - 1}{n}}\beta_{n-1}, \qquad n > 1.
\label{40}
\end{equation}
Приведем это представление, несколько упростив обозначения и заодно исправив
замеченную опечатку. С этой целью введем обозначения

${\bf M}(n) = \{{\bf m}\}$ --- множество $(n - 1)$-мерных векторов
${\bf m} = (m_1, m_2, \ldots, m_{n-1})$, ком\-по\-нен\-ты которых являются
целыми
неотрицательными числами, удовлетворяющими условию:
\begin{equation}
\sum_{j=1}^{n - 1}jm_j = n - 1.
\label{41}
\end{equation}

Для каждого вектора ${\bf m} \in {\bf M}(n)$ определим {\bf норму вектора},
обозначив ее $||{\bf m}||$ и полагая
\begin{equation}
||{\bf m}|| = \sum_{i = 1}^{n - 1} m_i.
\label{42}
\end{equation}

В этих обозначениях величина $\beta_{\mu}$ представляется следующим образом:
\begin{equation}
\beta_{\mu} = -{\frac{1}{{\mu}!}}
\sum_{{\bf m} \in {\bf M}({\mu} + 1)}
({\mu} + ||{\bf m}|| - 1)!
\prod_{j = 1}^{\mu}\frac{1}{m_j!}(-(j + 1)b_{j + 1})^{m_j}.
\label{43}
\end{equation}

Из формул (40) и (43) вытекают представления вириальных коэффициентов
в виде многочленов от майеровских коэффициентов $b_n$:
\begin{equation}
B_n = -\frac{n - 1}{n!}\sum_{{\bf m} \in {\bf M}(n)}(n + ||{\bf m}|| - 2)!
\prod_{j = 1}^{n-1}\frac{1}{m_j!}(-(j + 1)b_{j + 1}))^{m_j}.
\label{44}
\end{equation}
Формулу (44) будем называть {\bf формулой Майера}.

Пусть в формуле (44)  майеровские коэффициенты $b_n$ определяются
по формулам (32) и (30) их представлениями в виде древесных
сумм. Тогда формулы (44), (32) и (30) являют собой
представления вириального коэффициента $B_n$ в виде многочленов от
древесных сумм, представляющих майеровские коэффициенты
$b_2, b_3, \ldots, b_n$. Такое представление
вириального коэффициента $B_n$ будем называть его представлением
формулой Майера и формулами (32) и (30).

Для краткости сложность процедуры вычисления оценки вириального
коэффициента $B_n$, основанной на его представлении формулой Майера и
формулами (32) и (30) будем называть {\bf сложностью
представления вириального коэффициента формулой Майера и формулами
(32) и (30).}

Представляет интерес вопрос: какова же сложность вычислений оценок
вириальных коэффициентов с помощью этих представлений? Чтобы ответить
на этот вопрос, нужно четко определить процесс вычислений этих оценок.
В данной статье предлагается следующая схема этого процесса:

{\bf Этап 1.} Вычисление оценок майеровских коэффициентов, входящих в
представление данного вириального коэффициента по формуле Майера.

{\bf Этап 2.} Вычисление оценки данного вириального коэффициента.
Вычисление производится по формуле Майера, в которую вместо майеровских
коэффициентов под\-став\-ля\-ют\-ся вычисленные оценки этих коэффициентов.

Для оценки сложности этих вычислений по формуле Майера представим эту
фор\-му\-лу в несколько ином, более удобном для решения этой задачи виде.

С этой целью введем обозначения:

\begin{equation}
Q_n({\bf x}; {\bf y}; {\bf m}) =
\prod_{j = 1}^{n-1} \frac{1}{m_j!}(y_j x_j)^{m_j}.
\label{45}
\end{equation}
Здесь\quad ${\bf x} = (x_1, x_2, \ldots , x_{n - 1})$, \quad
${\bf y} = (y_1, y_2, \ldots , y_{n - 1})$ \quad
и\quad ${\bf m} = (m_1, m_2, \ldots , m_{n - 1})$ ---
\mbox{$(n - 1)$-мер}\-ные
векторы, причем компоненты вектора ${\bf m}$  являются целыми
не\-от\-ри\-ца\-тель\-ны\-ми
числами, удовлетворяющими условию (41).

Положим
\begin{equation}
x_i = -b_{i + 1}, \quad i = 1, 2, \ldots , n - 1; \quad
{\bf b} = \{b_2, b_3, \ldots, b_n\};
\label{46}
\end{equation}
\begin{equation}
y_i = i + 1, \quad i = 1, 2, \ldots , n - 1.
\label{47}
\end{equation}

В этих обозначениях формула Майера (44)
принимает вид

\begin{equation}
B_n =
-\frac{n - 1}{n!}\sum_{{\bf m} \in {\bf M}(n)}
(n + ||{\bf m}|| - 2)! Q_n({\bf x}; {\bf y}; {\bf m}).
\label{48}
\end{equation}

Из условия (41) вытекает, что норма любого вектора
$m \in {\bf M}(n)$ удовлетворяет неравенству
\begin{equation}
||{\bf m}|| \le n - 1.
\label{49}
\end{equation}

{\bf Замечание 10.}  Из определения функции
$Q_n({\bf x}; {\bf y}; {\bf m})$ формулой (45) следует,
что в случае, когда значения компонент вектора $\bf y$ вычисляются
по формулам (47), а  значения компонент вектора $\bf x$ заданы,
для вычисления значения функции $Q_n({\bf x}; {\bf y}; {\bf m})$
требуется произвести не более $5||{\bf m}||$ арифметических операций,

Также и в случае, когда значения компонент вектора $\bf y$ вычисляются
по формуле
\begin{equation}
y_i = -i - 1, \quad i = 1, 2, \ldots , n - 1, \quad
\label{47'}
\end{equation}
а  значения компонент вектора $\bf x$ заданы,
для вычисления значения функции $Q_n({\bf x}; {\bf y}; {\bf m})$
требуется произвести не более $5||{\bf m}||$ арифметических операций,

В случае же, когда все компоненты вектора $\bf y$ равны числу $1$,
а  значения компонент вектора $\bf x$ заданы,
для вычисления значения функции $Q_n({\bf x}; {\bf y}; {\bf m})$
при данных значениях векторов $\bf x$ и $\bf y$
требуется произвести не более $3||{\bf m}||$ арифметических операций.
$\blacksquare$

Из определения суммы $\sum\limits_{{\bf m} \in {\bf M}(n)}$
\rule[-3ex]{0em}{3ex}
следует, что число слагаемых в этой сумме равно числу всех неупорядоченных
разложений числа $n - 1$ в сумму натуральных слагаемых. Следуя [26, 29],
обозначим это число $p(n - 1)$.

{\bf Замечание 11.}  Величина $p(n)$ растет с ростом $n$ довольно
медленно. Ее значения приведены в книге [26, 29] (см. Таблицу 4.2).
Так, при $n = 9$ эта величина принимает значение $30$, а при $n = 10$
эта величина равна $42$. $\blacksquare$

Из замечания 11, из формулы (45) и из неравенства (49)
вытекает, что при $n \le 10$ для вычисления суммы
$$
 \sum_{{\bf m} \in {\bf M}(n)} (n + ||{\bf m}|| - 2)!
Q_n({\bf x}; {\bf y}; {\bf m}),
$$
где $\bf x$ и $\bf y$ определяются формулами (46) и (47)
соответственно, требуется произвести менее $2430$ арифметических
операций. Из этой оценки и формулы Майера следует, что для вычисления
по формуле Майера оценки вириального коэффициента $B_n$ по известным
оценкам майеровских коэффициентов $b_n$ при $n \le 10$ тре\-бу\-ет\-ся
произвести менее $2440$ арифметических операций.

Это --- ничтожно малое число арифметических операций по сравнению
с числом операций, необходимых для получения оценки даже первых
вириальных коэффициентов, таких как $B_4$, $B_5$, $B_6$ (и как
майеровских коэффициентов  $b_4$, $b_5$, $b_6$).
Ведь при процедуре вычисления оценок этих коэффициентов методом
Монте-Карло про\-из\-во\-дит\-ся порядка $10^{10}$ и более
статистических испытаний. Отсюда вытекает следующее

{\bf Замечание 12.} Сложность процесса вычисления оценки
вириального коэффициента $B_n$, основанного на
представлении этого коэффициента формулой Майера и формулами (32)
и (30), незначительно превышает
сложность всех вычислений, осуществляемых на первом этапе этого
процесса. Это дает возможность использовать критерий сложности всех
вычислений, осуществляемых на первом этапе,
в качестве критерия сложности представления этого вириального коэффициента
формулой Майера и формулами (32) и (30). $\blacksquare$

Разумеется, сложность процедуры вычисления оценок майеровских
коэффициентов зависит от их представлений. Для краткости сложность
процедуры вычисления оценок всех майеровских коэффициентов
из совокупности $\{b_2, b_3, \ldots, b_n\}$ с помощью данных
их представлений мы будем называть {\bf сложностью данной совокупности
представлений майеровских коэффициентов}.

Опираясь на замечание 12, будем считать, что критерием  сложности
представления вириального коэффициента $B_n$ формулой Майера
и формулами (32) и (30) является критерий сложности
совокупности, состоящей из представлений
майеровских коэффициентов $b_2, b_3, \ldots, b_n$.

В рассматриваемых ниже случаях майеровские коэффициенты представляются
формулами вида (32) и (30). По лемме 3 эти представления
являются базовыми линейными комбинациями с коэффициентами незначительной
сложности.

В дальнейшем изложении мы будем рассматривать только базовые линейные
комбинации интегралов с коэффициентами незначительной сложности. Для
краткости мы их будем называть {\bf базовыми линейными комбинациями}.

Для того, чтобы оценить сложность совокупности базовых
линейных комбинаций, представляющих майеровские коэффициенты
$b_2, b_3, \ldots, b_n$, нужно ввести критерии сложности оценки конечной
совокупности базовых линейных комбинаций с коэффициентами незначительной
сложности.

В дальнейшем изложении мы будем рассматривать только конечные
совокупности базовых линейных комбинаций интегралов с коэффициентами
незначительной сложности. Для краткости мы их будем называть
{\bf совокупностями базовыхми линейных комбинаций}.

Каждый из предлагаемых в статье критериев сложности оценки конечной
совокупности базовых линейных комбинаций порождается одним из изложенных
выше критериев сложности базовых линейных комбинаций. Обозначим $Cr'_i$,
где $i= 1, 2, 3$, критерий сложности оценки конечной совокупности базовых
линейных комбинаций, порожденный критерием $Cr_i$.

Для каждой конечной совокупности
${\mathfrak L}$ базовых линейных комбинаций критерий ее сложности
$Cr'_i({\mathfrak L})$, где $i = 1, 2, 3$ определяется формулой
\begin{equation}
Cr_i'({\mathfrak L}) = \sum_{L \in {\mathfrak L}} Cr_i(L), \quad
i = 1, 2,3.
\label{50}
\end{equation}

Пусть $L_0$ --- базовая линейная комбинация с коэффициентами
незначительной сложности. Поставим ей в соответствие
совокупность ${\mathfrak L}_0 = \{ L_0\}$, состоящую из одной линейной
комбинации $L_0$. Очевидно, что базовая линейная комбинация $L_0$ и
совокупность ${\mathfrak L}_0$ имеют одну и ту же сложность вычислений.

Совокупность ${\mathfrak L}_0$ по ее определению принадлежит множеству
конечных совокупностей линейных комбинаций с коэффициентами
незначительной сложности. Поэтому для нее при всех $i = 1, 2, 3$ 
определено значение критерия сложности $Cr'_i$.
По определению критерия $Cr'_i$ формулой (51) имеет место равенство
\begin{equation}
Cr'_i({\mathfrak L_0}) = Cr_i(L_0).
\label{51}
\end{equation}

Это равенство дает возможность сравнивать сложность данной базовой
линейной комбинации $L_0$  со сложностью конечной совокупности
${\mathfrak L'} = \{ L \}$ базовых линейных комбинаций
с коэффициентами незначительной сложности.

Используя равенство (52), дадим следующее

О\,п\,р\,е\,д\,е\,л\,е\,н\,и\,е 22. Пусть $L$ ---  базовая линейная
комбинация с коэффициентами незначительной сложности, а
${\mathfrak L}$ --- конечная совокупность базовых линейных комбинаций.
Будем считать, что {\bf по критерию $Cr'_i$, где $i = 1, 2, 3$,
базовая линейная комбинация $L$ значительно сложнее
совокупности базовых линейных комбинаций} ${\mathfrak L}$, если
$Cr_i(L) > Cr_i'({\mathfrak L})$. Если же 
$Cr_i(L) < Cr_i'({\mathfrak L})$, то будем
считать, что {\bf по критерию $Cr'_i$ базовая линейная комбинация $L$
значительно проще совокупности базовых линейных комбинаций}
${\mathfrak L}$.

В случае, когда $Cr_i(L) = Cr_i'({\mathfrak L})$, будем считать, что
{\bf по критерию $Cr'_i$ сложность базовой линейной комбинации $L$
приблизительно равна сложности совокупности базовых линейных
комбинаций} ${\mathfrak L}$, $\blacksquare$

{\bf Пример 4} В таблице 4 приведены вычисленные значения критерия
$Cr'_1({\mathfrak L}_{TR}(n))$ при \mbox{$n = \overline{2,10}$}, где 
${\mathfrak L}_{TR}(n)$ --- представление вириального коэффициента $B_n$
по формуле (44) Майера и формулам (32) и (30). 
В частности, $Cr'_1({\mathfrak L}_{TR}(8)) = 857$, 
$Cr_1({\mathfrak L}_{TR}(9)) = 3709$, $Cr'_1({\mathfrak L}(10)) = 17056$. 
Срав\-ни\-вая эти значения со значениями критерия сложности $Cr_1$ 
представлений Ри-Гувера, приведенными в таблице 1, мы видим, что значения
критерия $Cr_1({\mathfrak L}_{TR}(n))$ при $n = 8, 9, 10$ меньше значений
критерия сложности $Cr_1(L)$ для соответствующих представлений Ри-Гувера.
Стало быть, при этих значениях $n$ представление вириального коэффициента
$B_n$ формулами (44), (32) и (30) значительно проще
представления Ри-Гувера этого коэффициента. $\blacktriangleright$

{\bf Пример 5} Сравним по критерию $Cr'_3$ сложность представлений
вириальных коэффициентов $B_6$ и $B_7$ по методу Ри-Гувера со сложностью
их представлений в виде многочлена от древесных сумм по формулам
(44), (32) и (30). В таблице 6 приведены, в~частности,
следующие результаты:
\begin{equation}
Cr'_3({\mathfrak L}_{TR}(6)) = 228, \quad
Cr'_3({\mathfrak L}_{TR}(7)) < 2247, \quad
Cr_3( L_{RH}(6)) = 230, \quad Cr_3(L_{RH}(7)) = 2565.
\end{equation}

Из сравнения этих значений критериев сложности $Cr'_3$ и $Cr_3$ следует,
что при зна\-че\-ни\-ях $n = 6, 7$ представление вириального коэффициента
$B_n$ формулами (44), (32) и (30) проще представления
Ри-Гувера этого коэффициента. $\blacktriangleright$

Были также вычислены значения критерия $Cr'_2$ для совокупностей множеств вида
${\mathfrak L}(n)$ при $n = 2, 3, 4, 5, 6$. Результаты вычислений приведены
в таблице 5.

9. Перейдем теперь к представлениям вириальных коэффициентов в виде 
многочленов от древесных сумм, представляющих по формулам (38) и (30)
коэффициенты $a_n$.
Эти представления вириальных коэффициентов при $n > 1$ имеют 
вид [10, 11, 17, 36]:
\begin{equation}
B_n = \sum_{{\bf m} \in {\bf M}(n + 1)}||{\bf m}||!
e_{\scriptscriptstyle ||{\bf m}||}
\prod_{j=1}^n[\tau_j]^{m_j}(m_j!)^{-1}, \quad n \ge 2,
\label{53}
\end{equation}
где коэффициенты $e_{\mu}$ и $\tau_{\mu}$ определяются формулами
\begin{equation}
e_1 = \tau_1 = 1;\quad
e_{\mu} = \mu^{-1}
\sum_{{\bf m} \in {\bf M}(\mu)}||{\bf m}||!
\prod_{j=1}^{\mu - 1}(m_j!)^{-1}[(j + 1)a_{j + 1}]^{m_j},\quad \mu \ge 2;
\label{54}
\end{equation}

\begin{eqnarray}
\tau_{\mu} = (\mu - 1)!
\sum_{{\bf m} \in {\bf M}(\mu)}\left[(\mu-||{\bf m}||)!\,\right]^{-1}
\prod_{j = 1}^{\mu - 1}(m_j!)^{-1}\{-(j + 1)a_{j + 1}\}^{m_j}.
\quad \mu \ge 2.
\label{55}
\end{eqnarray}
По этим формулам вириальный коэффициент $B_n$
представляется в виде многочлена от древесных сумм, представляющих
коэффициенты $a_n$.

Представляет интерес вопрос: какова же сложность вычислений оценки
вириального коэффициента $B_n$ с помощью его представления формулами
(54), (55) и (56)?

Чтобы оценить сложность этих вычислений, прежде всего
представим вириальный коэффициент $B_n$ и величины $e_m$ и $\tau_m$
в более удобном для этой цели виде.

А именно, используя введенную формулой (45) функцию
$Q_n({\bf x}; {\bf y}; {\bf m})$,
 представления величин $B_n$, $e_m$ и $\tau_m$ формулами соответственно
(54), (55) и (56) преобразуем следующим образом:
\begin{equation}
e_1 = 1;\quad
e_{\mu} =
\mu^{-1}\sum_{{\bf m} \in {\bf M}(\mu)}
||{\bf m}||!Q_m({\bf x}; {\bf y}; {\bf m}),\quad \mu \ge 2,
\label{56}
\end{equation}
где
\begin{equation}
x_j = a_{j + 1},\quad y_j = j + 1, 1 \le j < \mu;
\label{57}
\end{equation}
\begin{eqnarray}
\tau_1 = 1; \quad \tau_{\mu} = (\mu - 1)!
\sum_{{\bf m} \in {\bf M}(\mu)}
\left\{[\mu - ||{\bf m}||]!\,\right\}^{-1}
Q_m({\bf x}; -{\bf y}; {\bf m}),   \quad \mu \ge 2,
\label{58}
\end{eqnarray}
где векторы ${\bf y}$ и ${\bf x}$  определены формулами (58), а
вектор $-{\bf y}$ определен формулой
\begin{equation}
-{\bf y} = (-y_1, -y_2, \ldots, -y_{\mu - 1});
\label{59}
\end{equation}
\begin{equation}
B_n = \sum_{{\bf m} \in {\bf M}(n + 1)}
||{\bf m}||!\,{\displaystyle e}_{\scriptscriptstyle ||{\bf m}||}
Q_{n + 1}({\bf x}; {\bf y}; {\bf m}), \quad n \ge 2,
\label{60}
\end{equation}
где величины $e_j$ при $j = \overline{1.n}$ определены формулами
(57),
\begin{equation}
x_j = \tau_j, \quad y_j = 1 \quad \mbox{\rm при}\quad j = \overline{1,n},
\label{61}
\end{equation}
а величины $\tau_j$ определены формулами (59), где векторы ${\bf y}$ 
и ${\bf x}$  определены формулами (58), а вектор $-{\bf y}$ 
определен формулой (60).

В этих преобразованных представлениях вириальный коэффициент $B_n$ также,
как и в представлениях формулами (54), (55) и (56),
представляется в виде
многочлена от древесных сумм, представляющих коэффициенты $a_n$.

Далее, чтобы ответить на поставленный вопрос, нужно четко определить
процесс вычислений оценки вириального коэффициента $B_n$.
В данной статье предлагается следующая схема этого процесса:

{\bf Этап 1.} Вычисление оценок коэффициентов $a_k$ при всех
$k = \overline{2,n}$.

{\bf Этап 2.} Вычисление оценок совокупности величин
${\bf e}_n = \{e_2, e_3, \ldots, e_n\}$.
Вычисление производится по формуле (57), в которую вместо
коэффициентов $a_k$ под\-став\-ля\-ют\-ся вычисленные оценки этих
коэффициентов.

{\bf Этап 3.} Вычисление оценок совокупности величин
${\bf \bar \tau_n} = \{\tau_2, \tau_3, \ldots, \tau_n\}$.
Вычисление производится по формуле (59), в которую вместо
коэффициентов $a_k$ под\-став\-ля\-ют\-ся вычисленные оценки этих
коэффициентов.

{\bf Этап 4.} Вычисление оценки данного вириального коэффициента.
Вычисление производится по формуле (61), в которую вместо величин
$e_{\mu}$ и $\tau_{\mu}$ под\-став\-ля\-ют\-ся вычисленные оценки этих
величин.

Нашей ближайшей целью является оценить сверху число арифметических
операций, потребных для вычислений, осущетвляемых на этапах 2--4.

Введем обозначения:

\begin{equation*}
{\bf e}_n = (e_1, e_2,\ldots, e_n), \quad
\mbox{$\bm\tau_n$} = (\tau_1, \tau_2, \ldots, \tau_n), \quad
{\bf a}_n = \{a_1, a_2, \ldots, a_n\}, \quad n \ge 2;
\end{equation*}
$E_1(\mu, {\bf m}\,\mid {\bf a_n })$ --- оценка сверху числа
арифметических операций, которые при данном значении $\mu \le n$ и
данном векторе ${\bf m} \in {\bf M}(\mu)$ потребны для вычисления значения
произведения $||{\bf m}||!Q_{\mu}({\bf x}; {\bf y}; {\bf m})$, где
$(\mu - 1)$-мерные векторы  $\bf x$ и $\bf y$ определяются по формулам
(58), а множество ${\bf a_n }$ задано;{} \\ 
\mbox{$E_2(\mu, {\bf m} \mid {\bf a}_n)$} --- оценка сверху числа арифметических
операций, которые при данном значении $\mu \le n$ и
дан\-ном векторе ${\bf m} \in {\bf M}(\mu)$ потребны для вычисления
значения произведения
$\left\{[\mu - ||{\bf m}||]!\,\right\}^{-1}Q_{\mu}({\bf x};
-{\bf y}; {\bf m})$,
где $(\mu - 1)$-мерные векторы ${\bf x}$ и ${\bf y}$  определяются
формулами (58), вектор ${-\bf y}$ определяется формулой (60), а
множество ${\bf a}_n$ задано;
\begin{equation}
\alpha(n, {\bf m}\mid {\bf e}_n, \mbox{$\bm\tau_n$}) =
||{\bf m}||!\,{\displaystyle e}_{\scriptscriptstyle ||{\bf m}||}
Q_{n + 1}({\bf x}; {\bf y}; {\bf m}), \quad {\bf m} \in {\bf M}(n + 1),
\label{62}
\end{equation}
где $n$-мерные векторы ${\bf y}$ и $\bf x$ определяются
по формулам (62), а множества ${\bf e}_n$ и \mbox{$\bm\tau_n$}
заданы;

$E_3(n, {\bf m}\mid {\bf e}_n, \mbox{$\bm\tau_n$})$ --- оценка числа
арифметических операций, которые при данном век\-то\-ре
${\bf m} \in {\bf M}(n + 1)$ потребны для вычисления значения
произведения
\mbox{$\alpha({\bf m}\mid {\bf e}_n, \mbox{$\bm\tau_n$})$}, где
$n$-мерные векторы ${\bf y}$ и $\bf x$ определяются по формулам
(62),
а множества $\bf e_n$ и \mbox{$\bm\tau_n$} заданы;

$E(e_{\mu}\mid {\bf a}_{\mu})$ --- оценка сверху числа арифметических
операций, потребных для вычисления значения величины $e_{\mu}$ при данных
значениях множества коэффициентов
${\bf a}_{\mu} = \{a_1, a_2, \ldots, a_{\mu}\}$;

$E(\tau_{\mu} \mid {\bf a}_{\mu})$ --- оценка сверху числа арифметических
операций, потребных для вычисления значения величины $\tau_{\mu}$
при данных значениях множества
коэффициентов ${\bf a}_{\mu} = \{a_1, a_2, \ldots, a_{\mu}\}$;

$E({\bf e}_n \mid {\bf a}_n)$ --- оценка сверху числа арифметических операций,
потребных для вычисления значения совокупности величин
${\bf e}_n = \{e_1, e_2, \ldots, e_n\}$ при данных значениях множества
коэффициентов ${\bf a}_n$;

$E(\mbox{$\bm\tau_n$} \mid {\bf a}_n)$ --- оценка сверху числа 
арифметических операций, потребных для вычисления значения совокупности 
величин $\mbox{$\bm\tau_n$} = \{\tau_1, \tau_2, \ldots, \tau_n\}$ 
при данных значениях множества коэффициентов 
$({\bf a})_n = \{a_1, a_2, \ldots, a_n\}$;

$E(B_n\mid\, {\bf e}_n, \mbox{$\bm\tau_n$})$ --- оценка сверху числа
арифметических операций, потребных для вычисления значения вириального
коэффициента $B_n$ при данных значениях совокупности величин
${\bf e}_n = \{e_1, e_2, \ldots, e_n\}$ и совокупности величин
$\mbox{$\bm\tau_n$} = \{\tau_1, \tau_2, \ldots, \tau_n\}$;

$E(B_n\mid\, {\bf a}_n)$ --- оценка сверху числа арифметических операций,
потребных для вычисления значения вириального коэффициента $B_n$
 при данных оценках множества коэффициентов
${\bf a}_n = \{a_1, a_2, \ldots, a_n\}$.

Найдем оценку сверху числа арифметических операций, потребных
для вычисления оценок совокупности величин
${\bf e}_n = \{e_1, e_2, \ldots, e_n\}$ при данных значениях множества
коэффициентов ${\bf a}_n$.

Из определения множества векторов ${\bf M}(\mu)$ следует, что при любом
${\mu} \ge 2$ всякий вектор ${\bf m} \in {\bf M}(\mu)$ удовлетворяет
неравенству
\begin{equation}
||{\bf m}|| \le \mu - 1.
\label{64}
\end{equation}

Из определения оценки $E_1(\mu, {\bf m} \mid {\bf a}_n)$, 
определения функции $Q_n({\bf x}; {\bf y};{\bf m})$ формулой (45), 
неравенства (64)  и замечания 9 вытекает, что при любом 
${\mu} \ge 2$ и любом векторе ${\bf m} \in {\bf M}(\mu)$ имеет место  
неравенство
\begin{equation}
E_1(\mu, {\bf m} \mid {\bf a}_n\,) \le 7(\mu - 1).
\label{65}
\end{equation}

Из определения величины $e_{\mu}$ формулой (57), неравенства (65),
замечания 10 и определений оценок $E(e_{\mu}\mid {\bf a}_{\mu})$ и
$E_1(\mu, {\bf m} \mid {\bf a}_n)$ вытекает оценка
\begin{equation}
E(e_{\mu} \mid {\bf a}_{\mu}) =
\sum_{{\bf m} \in {\bf M}(\mu)}E_1(\mu, {\bf m}|\,{\bf a}_{\mu}) \le
7p(\mu -1)(\mu - 1).
\label{66}
\end{equation}

Используя неравенство (66) и монотонное возрастание функции $p(n)$,
из определений оценок $E(e_{\mu} \mid {\bf a}_{\mu})$ и
$E({\bf e}_n \mid {\bf a}_n)$ получаем неравенство
\begin{equation}
E({\bf e}_n \mid {\bf a}_n) =
\sum_{\mu = 2}^n E(e_{\mu} \mid {\bf a}_{\mu}) \le
7p(n - 1)\sum_{\mu = 2}^n (\mu - 1) = 7p(n - 1)n(n - 1)/2.
\label{67}
\end{equation}

Найдем оценку сверху числа арифметических операций, потребных
для вычисления оценок совокупности величин
$\mbox{$\bm\tau_n$} = \{\tau_1, \tau_2, \ldots, \tau_n\}$ при данных оценках
множества коэффициентов ${\bf a}_n$.

Из определения оценки $E_2(\mu, {\bf m}\mid {\bf a}_n)$,
определения функции $Q_n({\bf x}; {\bf y};{\bf m})$ формулой (45),
неравенства (64) и замечания 9 вытекает, что при любом ${\mu} \ge 2$
и любом векторе ${\bf m} \in {\bf M}(\mu)$ имеет место  неравенство
\begin{equation}
E_2(\mu, {\bf m}|\,{\bf a}_n) \le 7(\mu - 1).
\label{68}
\end{equation}

Из определения величины $\tau_{\mu}$ формулой (59), неравенства
(68), замечания 10 и определений оценок 
$E(\tau_{\mu} \mid {\bf a}_{\mu})$ и
$E_2(\mu, {\bf m} \mid {\bf a}_n)$ вытекает оценка
\begin{equation}
E(\tau_{\mu} \mid {\bf a}_n) =
\sum_{{\bf m} \in {\bf M}(\mu)}E_2(\mu, {\bf m}|\,{\bf a}_n) \le
7p(\mu -1)(\mu - 1).
\label{69}
\end{equation}

Используя неравенство (69) и монотонное возрастание функции $p(n)$,
из определений оценок $E(\tau_{\mu} \mid {\bf a}_n)$ и
$E(\mbox{$\bm\tau_n$} \mid {\bf a}_n)$ получаем
неравенство
\begin{equation}
E(\mbox{$\bm\tau_n$} \mid {\bf a}_n) =
\sum_{\mu = 1}^n E(\tau_{\mu} \mid {\bf a}_n) \le
7p(n - 1)\sum_{\mu = 2}^n (\mu - 1) =
7p(n - 1)n(n - 1)/2.
\label{70}
\end{equation}

Найдем оценку сверху числа арифметических операций, потребных
для вычисления оценки данного вириального коэффициента $B_n$ при
данных оценках совокупности величин
${\bf e}_n = \{e_1, e_2, \ldots, e_n\}$ и данных оценках
 совокупности величин
$\mbox{$\bm\tau_n$} = \{\tau_1, \tau_2, \ldots, \tau_n\}$.

Из неравенства (64), определения произведения
$\alpha(n, {\bf m} \mid {\bf e}_n, \mbox{$\bm\tau_n$})$
формулой (63), определения оценки
$E_3(n, {\bf m} \mid {\bf e}_n, \mbox{$\bm\tau_n$})$,
определения функции $Q_n({\bf x}; {\bf y};{\bf m})$
формулой (45) и замечания 9 вытекает, что при любом $n \ge 2$
и любом векторе ${\bf m} \in {\bf M}(n + 1)$ имеет место  неравенство
\begin{equation}
E_3(n, {\bf m} \mid {\bf e}_n, \mbox{$\bm\tau_n$}) \le 5n.
\label{71}
\end{equation}

 Из определения вириального коэффициента $B_n$ формулой (61) и
определения произведения 
\mbox{$\alpha(n, {\bf m} \mid {\bf e}_n, \mbox{$\bm\tau_n$})$}
формулой (63) следует, что коэффициент $B_n$ может быть представлен
суммой
\begin{equation}
B_n =
\sum_{{\bf m} \in {\bf M}(n + 1)}
\alpha(n, {\bf m} \mid {\bf e}_n, \mbox{$\bm\tau_n$}).
\label{72}
\end{equation}

Отсюда, используя определения оценок 
$E_3(n, {\bf m} \mid {\bf a}_n)$ и
$E(B_n \mid {\bf e}_n, \mbox{$\bm\tau_n$})$, получаем неравенство
\begin{equation}
E(B_n \mid {\bf e}_n, \mbox{$\bm\tau_n$})) \le
\sum_{{\bf m} \in {\bf M}(n + 1)} E_3(n, {\bf m}\mid {\bf a}_n).
\label{73}
\end{equation}

Отсюда по замечанию 10 и неравенству (71) следует оценка
\begin{equation}
E(B_n \mid {\bf e}_n, \mbox{$\bm\tau_n$})) \le 5np(n).
\label{74}
\end{equation}

Из предложенной схемы процесса вычислений оценки вириального коэффициента
$B_n$ вытекает, что единственной целью всех вычислений
на этапах 2, 3 и 4 этой схемы является оценка этого коэффициента по
вычиcленным на этапе 1 оценкам коэффициентов $a_1, a_2, \ldots, a_n$.
Количество всех арифметических операций, потребных для достижения этой
цели есть сумма всех арифметических операций, которые следует осуществить
на этих этапах. Отсюда, применяя определения оценок
$E({\bf e}_n \mid {\bf a}_n)$, $E(\mbox{$\bm\tau_n$} \mid {\bf a}_n)$,
\mbox{$E(B_n \mid {\bf e}_n, \mbox{$\bm\tau_n$})$} и 
$E(B_n\mid {\bf a}_n)$,
получаем оценку
\begin{equation}
E(B_n \mid {\bf a}_n) = E({\bf e}_n \mid {\bf a}_n) +
E(\mbox{$\bm\tau_n$} \mid {\bf a}_n) +
E(B_n \mid {\bf e}_n, \mbox{$\bm\tau_n$})).
\label{75}
\end{equation}
Из прeдставления оценки $E(B_n \mid {\bf a}_n)$ формулой (75),
применяя неравенства (67), (70) и (74), получаем
\begin{multline}
E(B_n \mid {\bf a}_n) \le 7p(n - 1)n(n - 1)/2 + 7p(n - 1)n(n - 1)/2 +
p(n)5n ={}\\
7p(n - 1)n(n - 1) + 5np(n).
\label{76}
\end{multline}

В частности, при $n \le 10$ из формулы (76) и замечания 10 следует, что
для вычисления оценки вириального коэффициента $B_n$ по вычисленным
на этапе 1 оценкам коэффициентов $a_2, a_3, \ldots, a_n$ требуется
произвести менее $21000$ арифметических операций.

Это --- ничтожно малое число арифметических операций по сравнению с числом
операций, необходимых для получения оценки любого из коэффициентов
$a_4, a_5, \ldots$. Ведь при процедуре вычисления оценок этих коэффициентов
методом Монте-Карло производится порядка $10^{10}$ и более статистических
испытаний. Отсюда следует

{\bf Замечание 13.} Ос\-нов\-ная сложность процедуры вычисления оценки
вириального коэффициента по\-сред\-с\-т\-вом его представления
в виде многочлена от коэффициентов $a_n$ по формулам (54),
(55), (56), (38) и (30) при $n \ge 4$ состоит
в сложности процедуры вычисления оценок
всех  коэффициентов из совокупности $\{a_2, a_3, \ldots, a_n\}$.  При
этом сложность процедуры вычисления оценки вириального коэффициента $B_n$
незначительно превышает сложность процедуры вычисления оценок всех
коэффициентов $a_m$ из этой совокупности. Значит, критерий сложности 
представления этой совокупности является критерием сложности данного 
представления вириального коэффициента $B_n$. $\blacksquare$

По определению 21 и лемме 4 коэффициенты $a_n$ представляются 
по формулам (38) и (30) древесными суммами, которые являются базовыми
линейными комбинациями с коэффициентами незначительной сложности.
Поэтому критериями сложности оценки конечной совокупности этих
древесных сумм являются критерии
$Cr'_1({\mathfrak L})$, $Cr_2'({\mathfrak L})$
и  $Cr'_3({\mathfrak L})$ базовых линейных комбинаций. Совокупность
древесных
сумм, каждая из которых представляет по формулам (38) и (30)
коэффициент из совокупности коэффициентов $\{a_2, a_3, \ldots, a_n\}$
обозначим ${\mathfrak L}(n, 0) = \{L\}$.

В качестве примера при \mbox{$n = \overline{2,10}$} были вычислены 
значения критерия $Cr'_1({\mathfrak L})$ для совокупностей древесных сумм вида
${\mathfrak L}(n, 0) = \{L\}$.
Результаты приведены в таблице~4. Далее, при $n = \overline{2,6}$ были
вычислены значения критериев $Cr_2'({\mathfrak L})$ и $Cr'_3({\mathfrak L})$
для совокупностей древесных сумм  ${\mathfrak L}(n, 0) = \{L\}$.
Результаты приведены в таблицах 5 и 6 соответственно. Наконец, была
вычислена оценка сверху значения критерия $Cr'_3({\mathfrak L})$
для совокупности древесных сумм ${\mathfrak L}(7, 0) = \{L\}$.

Сравнение значений этих
критериев с приведенными в этих же таблицах значениями этих критериев
для представлений вириальных коэффициентов $B_n$ по методу Ри-Гувера
приводит к следующим выводам. Представление вириального коэффициента 
$B_n$ по методу Ри-Гувера значительно сложнее представления этого 
коэффициента многочленом от коэффициентов $a_n$, представленных 
древесными суммами по формулам (38) и (30), в следующих случаях: 
по критерию $Cr'_1({\mathfrak L})$ --- при $n = \overline{7,10}$, 
по критерию $Cr'_2$ --- при $n = 6$, по критерию $Cr'_3$ ---
при $n = 5, 6, 7$. Представление вириального коэффициента $B_4$
по методу Ри-Гувера значительно проще представления этого коэффициента
многочленом от коэффициентов $a_n$, представленных древесными суммами
по формулам (38) и (30). По критерию $Cr'_3$ при $n = 2, 3$
сложность представления вириального коэффициента $B_n$ по методу 
Ри-Гувера незначительно отличается от сложности представления 
многочленом от коэффициентов $a_n$, представленных древесными суммами 
по формулам (38) и (30).

10. Методом каркасных сумм можно получить и представления коэффициентов 
степенных рядов, не являющиеся древесными суммами. Так, методом 
каркасных сумм автором было получены представления вириальных 
коэффициентов в виде:
\begin{equation}
B_n = -\frac{n-1}{n!}\sum_{{\bf C}\in{\mathfrak C}(n)}J({\bf C}).
\label{77}
\end{equation}

\par

Здесь \mbox{${\mathfrak C}(n)$} --- множество ансамблей каркасных 
циклов [14--16, 18--20, 37, 38] всех двусвязных графов с~множеством 
вершин $V_n = \{1, 2, \ldots , n\}$;
$\bf C$ --- ансамбль каркасных циклов из множества ${\mathfrak C}(n)$;

\begin{eqnarray}
J({\bf C}) =
\int_{(\R)^{n-1}}\prod_{\{u,v\}\in X(S({\bf C}))}f_{uv}
\prod_{\{\widetilde u,\widetilde v\}\in X_{ad}({\bf C})}
(1+f_{\widetilde u, \widetilde v})(d{\bf r})_{1,n-1},
\label{78}
\end{eqnarray}

где
$S(\bf C)$ --- объединение всех циклов ансамбля $\bf C$
[14, 15, 19, 37];
$X(S({\bf C}))$ --- множество всех ребер графа $S(\bf C)$
[14, 15, 19, 37];
$X_{ad}({\bf C})$ --- множество всех допустимых ребер
[14, 15, 19, 37] ансамбля ${\bf C}$;
$\{u, v\}$ --- ребро, инцидентное вершинам $u$ и $v$.

Из определения интегралов вида $J({\bf C})$ формулой (78) следует,
что в каждом из интегралов, являющихся слагаемыми суммы в правой части
(77), подынтегральная функция пред\-ста\-в\-ля\-ет собой
произведение майеровских функций, помеченных ребрами циклов, входящих
в ансамбль каркасных циклов, маркирующий данный интеграл,
и больцмановских функций, помеченных ребрами из множества
$X_{ad}({\bf C}) = \{\{u,v\}\}$.
Такую сумму интегралов мы будем называть {\bf каркасной суммой.}

из определения  множества $X_{ad}({\bf C})$ следует.
что это множество состоит из попарно различных ребер, а каждое ребро,
содержащееся в этом множестве, соединяет две несмежные вершины
графа $S({\bf C})$.

{\bf Теорема 4.} {\it При любом ансамбле каркасных циклов
${\bf C}\in{\mathfrak C}(n)$ интеграл $J({\bf C})$ яв\-ля\-ет\-ся 
базовым интегралом порядка $n$.}

{\bf Доказательство.} Сначала докажем, что подынтегральная функция
интеграла $J({\bf C})$ является базовой.

С этой целью прежде всего докажем, что множества ребер $X(S({\bf C}))$ и
$X_{ad}({\bf C})$ образуют каноническую пару множеств
${\bf X} = (X(S({\bf C})), X_{ad}({\bf C}))$. Из определения множества
ребер $X(S({\bf C}))$ следует, что это множество состоит из попарно 
различных ребер. Как было отмечено выше, множество $X_{ad}({\bf C})$ 
также состоят из попарно различных ребер, а каждое ребро,
содержащееся в этом множестве, соединяет две несмежные вершины графа
$S({\bf C})$. Отсюда  следуют два вывода:

1) непересекающиеся множества $X(S({\bf C}))$ и $X_{ad}({\bf C})$
образуют упорядоченную пару ${\bf X} = (X(S({\bf C})), X_{ad}({\bf C}))$
 множеств;

2) вершины всех ребер из множества $X_{ad}({\bf C})$ принадлежат множеству
вершин графа $S({\bf C})$.

Так как $\bf C$ --- ансамбль каркасных циклов из
множества ${\mathfrak C}(n)$, то, как известно \cite{42},
граф $S({\bf C})$ является двусвязным графом с множеством
вершин $V_n$.

 Значит, имеет место равенство
\begin{equation}
V(X(S({\bf C}))) \cup V(X_{ad}({\bf C})) = V_n,
\label{79}
\end{equation}
где $V(X(S({\bf C})))$ --- множество всех вершин графа $S({\bf C})$, а
$V(X_{ad}({\bf C}))$ ---  множество вершин всех допустимых ребер 
ансамбля $\bf C$.

Из равенства (79) по определению 5 следует, что упорядоченная пара
множеств
${\bf X} = (X(S({\bf C}), X_{ad}({\bf C}))$ является канонической парой
порядка $n$. 

Из полученных результатов вытекает, что граф $S({\bf C})$ с 
поставленным ему в соответствие множеством $X_{ad}({\bf C})$ 
принадлежит множеству графов ${\widetilde {\mathfrak G}}_n$ 
по определению этого множества. 
Отсюда по теореме 1 следует, что определенное формулой (19) 
произведение $\widetilde P_{1n}(S({\bf C}))$ майеровских и 
больцмановских функций является базовым произведением порядка $n$.

В данном случае формула (19) имеет вид:
\begin{equation}
\widetilde P_{1n}(S({\bf C})) =
\prod_{\{i,j\} \in X(S({\bf C}))}\prod_{\{i',j'\} \in X_{ad}({\bf C})}
f_{ij}\widetilde f_{i'j'}.
\label{80}
\end{equation}
Из сравнения формул (78) и (80) следует, что подынтегральная
функция интеграла $J({\bf C})$ тождественна базовому произведению 
функций $\widetilde P_{1n}(S({\bf C}))$ и, следовательно, является 
базовым произведением функций порядка $n$.

Отсюда следует, что интеграл $J({\bf C})$ является, по определению 9, 
базовым интегралом порядка $n$. Теорема 4 доказана.
$\blacktriangleright$

Из теоремы 4 вытекает следующее

{\bf Следствие 6.} {\it Каркасная сумма в правой части {\rm (77)}
является,
по определению {\rm 11} и замечанию {\rm 5}, базовой линейной комбинацией с
коэффициентами незначительной сложности.}

Это обстоятельство позволяет использовать предлагаемые в этой статье
критерии $Cr_1$, $Cr_2$ и $Cr_3$ для сравнения по сложности 
представлений вириальных коэффициентов каркасными суммами с иными  
базовыми линейными комбинациями с коэффициентами незначительной 
сложности. Это обстоятельство также позволяет использовать
предлагаемые в этой статье критерии $Cr'_1$, $Cr_2'$ и $Cr'_1$
для сравнения по сложности представлений вириальных коэффициентов
каркасными суммами c многочленами от базовых линейных комбинаций
с коэффициентами незначительной сложности.

Из таблиц 1, 2, 3, 4, 5 и 6 вытекают следующие выводы.

По критериям $Cr_1$, $Cr_2$ и $Cr_3$ сложность представления 
вириального коэффициента $B_3$ каркасной суммой по формулам (77) и (78)
равна сложности представления Ри-Гувера этого вириального коэффициента 
и незначительно отличается от сложности представления коэффициента 
$a_3$ древесной суммой по формулам (38) и (30).

По критериям $Cr_1$ и $Cr_2$ это представление вириального коэффициента
$B_3$ каркасной суммой проще представления майеровского коэффициента
$b_3$ древесными суммами по формулам (32) и (30).
Но по критерию $Cr_3$ эти два представления по своей сложности
незначительно отличаются  друг от друга.

По критериям $Cr'_1$ и $Cr_2'$ представление вириального
коэффициента $B_3$ кар\-кас\-ной суммой проще его представления
формулой  (44) в виде многочлена от древесных сумм,
представляющих коэффициенты $b_n$ по формулам (32) и (30);
и проще его представления формулами (55), (56) и (57)
в виде многочлена от дре\-вес\-ных сумм, представляющих
коэффициенты $a_n$ по формулам (38) и (30).
Но по критерию $Cr'_3$ все эти три представления по своей сложности
незначительно отличаются друг от друга.

По критериям $Cr_1$, $Cr_2$ и $Cr_3$ представление вириального
коэффициента $B_4$ каркасной суммой по формулам (77) и (78)
сложнее представления Ри-Гувера этого вириального коэффициента и
сложнее представления коэффициента $a_4$ древесной
суммой по формулам (38) и (30).

Сложность представления вириального коэффициента $B_4$ каркасной
суммой по критерию $Cr_1$ не отличается от сложности представления
майеровского коэффициента $b_4$ древесной суммой по формулам (32)
и (30). Однако по критериям $Cr_2$ и $Cr_3$ это представление
вириального коэффициента $B_4$ сложнее вышеупомянутого представления
майеровского коэффициента $b_4$. Так как критерии $Cr_2$ и $Cr_3$
являются более точными, то, видимо, следует считать, что 
представление вириального коэффициента $B_4$ каркасной суммой сложнее
этого представления майеровского коэффициента $b_4$.

Наконец, представление вириального коэффициента $B_4$ каркасной суммой
по критериям $Cr'_1$ и $Cr_2'$ проще  представления этого коэффициента
формулой (44) в виде многочлена от дре\-весных сумм, 
пред\-став\-ля\-ю\-щих коэффициенты $b_n$ по формулам (32) и (30). Но
по критерию $Cr'_3$ первое из этих двух представлений вириального 
коэффициента $B_4$ сложнее второго.
Так как критерий $Cr'_3$ является более точным, чем критерии $Cr'_1$
и $Cr_2'$, то, видимо, следует считать, что данное
представление вириального коэффициента $B_4$ каркасной суммой сложнее
его представления в виде многочлена от древесных сумм, представляющих
коэффициенты $b_n$.

Автор считает своим приятным долгом выразить свою глубокую 
благодарность проф. Д.А. Кофке за быструю и эффективную информационную
поддержку и за~быст\-рые
и обстоятельные ответы на интересуюшие автора вопросы, проф. 
Г.А. Мартынову --- за эффективную информационную поддержку и полезные 
обсуждения, доктору Р.~Хеллману --- за~быструю и эффективную 
информационную поддержку, доктору Н.~Клисби --- за~быс\-трые 
и эф\-фек\-тив\-ные информационную и организационную под\-держ\-ки 
и высокую оценку работ автора
и к.ф.-м.н. В.И. Цебро --- за быстрые и эффективые информационную и
техническую под\-держ\-ки и полезные советы.

\pagebreak

\begin{center}
\bf  Таблицы cложности представлений древесными суммами майеровских
коэффициентов и коэффициентов $a_n$, представлений каркасными суммами вириальных
коэффициентов и представлений Ри-Гувера вириальных коэффициентов
\end{center}

\begin{center} \bf Таблица 1 cложности по критерию $Cr_1$
\end{center}
\begin{center}
\begin{tabular}{rccccccccc}
$n$ & 2 & 3 & 4 & 5 & 6 & 7 & 8 & 9 & 10 \\
$Cr_1(L_{TR}(n))$ & 1 & 2 & 5 & 14 & 44 & 157 & 634 & 2852 & 14047 \\
$Cr_1(L_{TR}(n.0))$ & 1 & 1 & 2 & 5 & 15 & 55 & 239 & 1169 & 6213 \\
$Cr_1(L_F(n))$ & 1 & 1 & 5 & 49 & 784 & - & - & - & - \\
$Cr_1(L_{RH}(n))$ & 1 & 1 & 2 & 5 & 23 & 171 & 2606 & 81564 & 4 980 756
\end{tabular}
\end{center}

\begin{center} \bf Таблица 2 cложности по критерию $Cr_2$
\end{center}
\begin{center}
\begin{tabular}{rccccccccc}
$n$ & 2 & 3 & 4 & 5 & 6 \\
$Cr_2(L_{TR}(n))$ & 1 & 5 & 23 & 93 & 403 \\
$Cr_2(L_{TR}(n,0))$ & 1 & 3 & 11 & 42 & 172 \\
$Cr_2(L_F(n))$ & 1 & 3 & 26 & - & - & \\
$Cr_2(L_{RH}(n))$ & 1 & 3 & 12 & 50 & 345 \\
\end{tabular}
\end{center}

\begin{center} \bf Таблица 3 cложности по критерию $Cr_3$
\end{center}
\begin{center}
\begin{tabular}{rccccccccc}
$n$ & 2 & 3 & 4 & 5 & 6 \\
$Cr_3(L_{TR}(n))$ & 0 & 1 & 8 & 37 & 183 \\
$Cr_3(TR(n,0))$ & 0 & 1 & 8 & 38 & 167 \\
$Cr_3(L_F(n))$ & 0 & 1 & 11 & - & - \\
$Cr_3(L_{RH}(n))$ & 0 & 1 & 6 & 30 & 230 \\
\end{tabular}
\end{center}

В таблицах приняты следующие обозначения:

$n$ --- номер майеровского(вириального)  коэффициента;

$L_{TR}(n)$ --- представление древесной суммой майеровского коэффициента
$b_n$, определяемого по формулам (32) и (30);

$L_{TR}(n.0)$  --- представление древесной суммой
коэффициента $a_n$, определяемого по формулам (38) и (30);

$L_F(n)$ --- представление каркасной суммой
вириального коэффициента $B_n$ по формулам (77) и (78);

$L_{RH}(n)$ --- представление вириального коэффициента $B_n$ по методу
Ри-Гувера;

\pagebreak

\begin{center}
\bf Таблицы  cложности представлений вириальных коэффициентов:
1)~представлений посредством майеровских коэффициентов, представленных
древесными суммами; 2) представлений посредством коэффициентов $a_n$,
представленных древесными суммами; 3) представлений каркасными суммами;
4) представлений Ри-Гувера;
\end{center}

\begin{center} \bf Таблица 4 cложности по критерию $Cr'_1$
\end{center}
\begin{center}
\begin{tabular}{rccccccccc}
$n$ & 2 & 3 & 4 & 5 & 6 & 7 & 8 & 9 & 10 \\
$Cr'_1({\mathfrak L}_{TR}(n))$ & 1 & 3 & 8 & 22 & 66 & 223 & 857 & 3709 & 17056 \\
$Cr'_1({\mathfrak L}_{TR}(n.0))$ & 1 & 2 & 4 & 9 & 24 & 79 & 318 & 1487 & 7700 \\
$Cr'_1(L_F(n))$ & 1 & 1 & 5 & 49 & 784 & - & - & - & - \\
$Cr'_1(L_{RH}(n))$ & 1 & 1 & 2 & 5 & 23 & 171 & 2606 & 81564 & 4 980 756
\end{tabular}
\end{center}

\begin{center} \bf Таблица 5 cложности по критерию $Cr_2'$
\end{center}
\begin{center}
\begin{tabular}{rccccccccc}
$n$ & 2 & 3 & 4 & 5 & 6 \\
$Cr_2'({\mathfrak L}_{TR}(n))$ & 1 & 6 & 28 & 121 & 524 \\
$Cr_2'({\mathfrak L}_{TR}(n,0))$ & 1 & 4 & 15 & 57 & 229 \\
$Cr_2'(L_F(n))$ & 1 & 3 & 26 & - & - & \\
$Cr_2'(L_{RH}(n))$ & 1 & 3 & 12 & 50 & 345 \\
\end{tabular}
\end{center}

\begin{center} \bf Таблица 6 cложности по критерию $Cr'_3$
\end{center}
\begin{center}
\begin{tabular}{rccccccccc}
$n$ & 2 & 3 & 4 & 5 & 6 & 7 \\
$Cr'_3({\mathfrak L}_{TR}(n))$ & 0 & 1 & 9 & 45 & 228 & < 2247 \\
$Cr'_3({\mathfrak L}_{TR}(n,0))$ & 0 & 1 & 7 & 28 & 125 & 612 \\
$Cr'_3(L_F(n))$ & 0 & 1 & 11 & - & - \\
$Cr'_3(L_{RH}(n))$ & 0 & 1 & 6 & 30 & 230 & 2565 \\
\end{tabular}
\end{center}

В таблицах приняты следующие обозначения:

$n$ --- номер майеровского(вириального)  коэффициента;

${\mathfrak L}_{TR}(n)$ --- представление вириального коэффициента $B_n$
по формуле (44) Майера многочленом от всех древесных сумм, являющихся
представлениями майеровских коэффициентов $b_2, b_3, \ldots, b_n$,
по формулам (32) и (30);

${\mathfrak L}_{TR}(n.0)$  --- представление вириального коэффициента $B_n$
по формулам (61), (57), (58), (59) и (62) многочленом от всех древесных
сумм, являющихся представлениями коэффициентов $a_2, a_3, \ldots, a_n$
по формулам (38) и (30);

${\mathfrak L}_F(n)$ --- представление каркасной суммой
вириального коэффициента $B_n$ по формулам (77) и (78);

$L_{RH}(n)$ --- представление вириального коэффициента $B_n$ по методу
Ри-Гувера;

{\bf Примечание.} В таблицах 1 и 4 значения длин базовых линейных
комбинаций, являющихся представлениями вириальных коэффициентов $B_n$
по методу Ри-Гувера, заимствованы из статьи [27]. Значения критериев
$Cr_2$ и $Cr_3$ для представлений вириальных коэффициентов $B_n$ по методу
Ри-Гувера вычислены, исходя из определения [45, 46, 47] этих
представлений и используя значения длин
базовых линейных комбинаций, приведенные в [27].
\renewcommand{\refname}{\large\bf Список литературы}

\end{document}